\documentclass[twocolumn,journal]{IEEEtran}
\usepackage[T1]{fontenc}
\usepackage[latin9]{inputenc}
\usepackage{array}
\usepackage{float}
\usepackage{multirow}
\usepackage{amsmath}
\usepackage{amsthm}
\usepackage{amssymb}
\usepackage{stackrel}
\usepackage{graphicx}
\usepackage[unicode=true,
 bookmarks=true,bookmarksnumbered=true,bookmarksopen=true,bookmarksopenlevel=1,
 breaklinks=false,pdfborder={0 0 0},pdfborderstyle={},backref=false,colorlinks=false]
 {hyperref}
\hypersetup{pdftitle={Your Title},
 pdfauthor={Your Name},
 pdfpagelayout=OneColumn, pdfnewwindow=true, pdfstartview=XYZ, plainpages=false}

\makeatletter

\providecommand{\tabularnewline}{\\}
\floatstyle{ruled}
\newfloat{algorithm}{tbp}{loa}
\providecommand{\algorithmname}{Algorithm}
\floatname{algorithm}{\protect\algorithmname}

\theoremstyle{plain}
\newtheorem{thm}{\protect\theoremname}
\theoremstyle{remark}
\newtheorem{rem}[thm]{\protect\remarkname}

\usepackage[caption=false,font=footnotesize]{subfig}
\usepackage{cite}
\usepackage{diagbox}
\usepackage{flushend}
\usepackage{titlesec}

\makeatother

\providecommand{\remarkname}{Remark}
\providecommand{\theoremname}{Theorem}

\begin{document}
\title{Two new algorithms for maximum likelihood estimation of sparse covariance
matrices with applications to graphical modeling}
\author{Ghania Fatima, Prabhu Babu, and Petre Stoica}
\maketitle
\begin{abstract}
In this paper, we propose two new algorithms for maximum-likelihood
estimation (MLE) of high dimensional sparse covariance matrices. Unlike
most of the state-of-the-art methods, which either use regularization
techniques or penalize the likelihood to impose sparsity, we solve
the MLE problem based on an estimated covariance graph. More specifically,
we propose a two-stage procedure: in the first stage, we determine
the sparsity pattern of the target covariance matrix (in other words
the marginal independence in the covariance graph under a Gaussian
graphical model) using the multiple hypothesis testing method of false
discovery rate (FDR), and in the second stage we use either a block
coordinate descent approach to estimate the non-zero values or a proximal
distance approach that penalizes the distance between the estimated
covariance graph and the target covariance matrix. Doing so gives
rise to two different methods, each with its own advantage: the coordinate
descent approach does not require tuning of any hyper-parameters,
whereas the proximal distance approach is computationally fast but
requires a careful tuning of the penalty parameter. Both methods are
effective even in cases where the number of observed samples is less
than the dimension of the data. For performance evaluation, we test
the proposed methods on both simulated and real-world data and show
that they provide more accurate estimates of the sparse covariance
matrix than two state-of-the-art methods.
\end{abstract}


\section{Introduction and Problem formulation}

The estimation of covariance matrices is an extensively studied research
problem in the field of multivariate analysis due to its pivotal role
in a wide variety of applications such as high-dimensional classification
\cite{witten2011penalized}, spectral analysis \cite{stoica2005spectral},
computational biology \cite{schafer2005shrinkage}, system identification
\cite{stoicasystemidentification}, radar and wireless communication
\cite{aubry2017geometric}, portfolio optimization \cite{ledoit2004honey},
asset allocation and risk assessment \cite{yang2015robust,deshmukh2020improved,senneret2016covariance}
and analysis of relationships between components in graphical models
\cite{toh2002inference,schafer2005empirical}. 

Let $\mathbf{Y}\triangleq\left[\mathbf{y}_{1},\mathbf{y}_{2},\ldots,\mathbf{y}_{n}\right]\in\mathbb{R}^{p\times n}$
be the data matrix consisting of $n$ independent and identically
distributed realizations of a $p-$variate Gaussian random variable
$\mathbf{y}\in\mathbb{R}^{p}$ with zero mean and covariance matrix
$\boldsymbol{\Sigma}$ $\left(\mathbf{y}\sim\mathcal{N}\left(0,\boldsymbol{\Sigma}\right)\right)$.
The joint distribution of the multivariate Gaussian data $\mathbf{Y}$
is:
\begin{equation}
p\left(\boldsymbol{\Sigma}\right)=\left(2\pi\right)^{-\frac{pn}{2}}\left(\det\left(\boldsymbol{\Sigma}\right)\right)^{-\frac{n}{2}}\exp\left(-\frac{1}{2}\stackrel[i=1]{n}{\sum}\mathbf{y}_{i}^{T}\boldsymbol{\Sigma}^{-1}\mathbf{y}_{i}\right)
\end{equation}
from which the log-likelihood of the data follows:
\begin{equation}
\mathcal{L}\left(\boldsymbol{\Sigma}\right)=-\frac{pn}{2}\log\left(2\pi\right)-\frac{n}{2}\log\det\left(\boldsymbol{\Sigma}\right)-\frac{n}{2}\textrm{Tr}\left(\boldsymbol{\Sigma}^{-1}\mathbf{S}\right),\label{eq:1}
\end{equation}
where $\mathbf{S}=\frac{1}{n}\mathbf{Y}\mathbf{Y}^{T}$ is the sample
covariance matrix (SCM). To estimate $\boldsymbol{\Sigma}$ from the
data $\mathbf{Y}$, the maximum likelihood estimation (MLE) problem
can be cast as the minimization of the negative log-likelihood: 
\begin{equation}
\underset{\boldsymbol{\Sigma}\succ0}{\min}f\left(\Sigma\right)\triangleq\log\det\left(\boldsymbol{\Sigma}\right)+\textrm{Tr}\left(\boldsymbol{\Sigma}^{-1}\mathbf{S}\right)\label{eq:2}
\end{equation}
When the number of samples is greater than or equal to the number
of variables $\left(n\geq p\right)$, the matrix $\mathbf{S}$ is
non-singular with probability one and the problem (\ref{eq:2}) has
a unique solution $\hat{\boldsymbol{\Sigma}}=\mathbf{S}$. However,
when $n<p$, $\mathbf{S}$ is singular. In addition, when $p$ is
large, $\mathbf{S}$ is rather noisy due to the accumulation of a
large number of errors. Thus, even when the sample size is greater
than but comparable to the dimension of the data, $\mathbf{S}$ is
a poor estimate of $\boldsymbol{\Sigma}$. 

The massive data surge in recent years has led to the availability
of high-dimensional data with limited sample sizes (for instance in
applications such as mobile networks, social networks, and computational
biology), thus making the problem of accurate covariance matrix estimation
a challenging task. Therefore, any existing prior information on the
covariance matrix must be exploited for its accurate estimation. Several
works available in the literature exploit the intrinsic structure
of the covariance matrix such as sparsity, which is a common occurrence
in high-dimensional settings since many of the variables (features)
are only weakly correlated and imposing sparsity in the estimation
problem enables the detection of important relationships. 

Motivated by the fact that in high dimensional settings many variables
are weakly correlated, we formulate our problem as the estimation
of a covariance matrix under the constraint that the target matrix
is sparse, i.e., many of its elements are zero. Hence, the optimization
problem can be written as:
\begin{equation}
\begin{aligned}\underset{\boldsymbol{\Sigma}\succ0}{\min}\; & f\left(\boldsymbol{\Sigma}\right)\\
\mathrm{s.t.\;} & \boldsymbol{\Sigma}\in\mathcal{C}
\end{aligned}
\label{eq:3}
\end{equation}
where $\mathcal{C}$ is a set of sparse symmetric matrices. The assumption
of sparsity is not only practical, but it also ensures that the MLE
exists for cases with $n<p$. 

\subsection{Graph motivation}

The methods based on sparse covariance matrix estimation are widely
used for inferring networks and graphs in high-dimensional applications.
The marginal independence between random variables is encoded using
a bi-directional graph where the random variables are identified by
the graph vertices and two random variables are said to be marginally
independent if they do not have an edge between them. Graphical models
are a useful tool for discovering the structure in high-dimensional
data, which is of significant interest in many applications such as
social networks \cite{banerjee2008model} and biology \cite{kuismin2017estimation}. 

\begin{figure}[tbh]
\begin{centering}
\includegraphics[width=8.4cm,height=3.8cm]{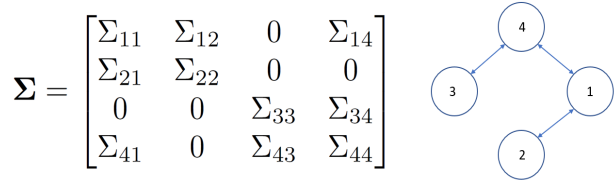}
\par\end{centering}
\caption{\label{fig:1-1}An example of a sparse covariance matrix with $p=4$
variables and its associated graph.}
\end{figure}

Gaussian graphical models for marginal independence, also known as
Gaussian covariance graph models \cite{cox1996multivariate}, impose
sparsity on the covariance matrix. The sparsity pattern of a covariance
matrix can be visualized with the help of covariance graph $\mathcal{G}\left(\mathcal{V},\mathcal{\mathcal{E}},\mathbb{I}\left(\boldsymbol{\Sigma}\right)\right)$
(notation explained at the end of this section), which has one vertex
for each of the variables and a bi-directional edge between two variables
if the covariance between them is non-zero. Fig. \ref{fig:1-1} shows
a sparse covariance matrix and its corresponding covariance graph.
The vertices $\mathcal{V}=\{1,2,3,4\}$ correspond to the indices
of the random variables $\left\{ y_{i}\right\} $. Two vertices $i$
and $j$ ($i\neq j$) are connected by a bidirectional edge $i\longleftrightarrow j$,
if $\Sigma_{ij}$ is non-zero, i.e., $\left\{ i,j\right\} \in\mathcal{E}$. 

\subsection{Literature review}

The most popular methods for sparse covariance estimation use regularized
estimation techniques such as banding \cite{wu2003nonparametric,bickel2008regularized},
tapering \cite{furrer2007estimation,cai2010optimal} and thresholding
\cite{el2008operator,rothman2009generalized,bickel2008covariance,cai2011adaptive}.
In applications where the natural ordering of variables is known and
the variables that are far apart are weakly correlated, methods based
on banding or tapering are employed. However, these methods are sensitive
to any permutation in the variables. In the absence of information
on the natural ordering of variables, element-wise thresholding, which
sets small entries of the SCM to zero, is useful. This is a simple
and direct method to induce sparsity which is permutation-invariant
with respect to indexing the variables. However, it does not guarantee
a positive definite estimate. Therefore, the selection of an appropriate
thresholding constant is important to balance the trade-off between
positive definiteness of the estimated matrix and the desired level
of sparsity. The authors of \cite{rothman2012positive} and \cite{xue2012positive}
resorted to the Frobenius norm approach with an $\ell_{1}$ norm penalty
to enforce sparsity (soft thresholding), in which the positive definiteness
of the estimate was ensured using a log-barrier term in the objective
function in \cite{rothman2012positive} and alternating directions
methods in \cite{xue2012positive}. 

Similar to penalizing the Frobenius norm, the likelihood in (\ref{eq:2})
can be penalized as follows:
\begin{equation}
\underset{\boldsymbol{\Sigma}\succ0}{\min}\;\log\det\left(\boldsymbol{\Sigma}\right)+\textrm{Tr}\left(\boldsymbol{\Sigma}^{-1}\mathbf{S}\right)+\lambda\left\Vert \mathbf{P}*\boldsymbol{\Sigma}\right\Vert _{1},\label{eq:5-1}
\end{equation}
where $\lambda$ is the penalty parameter and $\mathbf{P}$ is a weight
matrix (two common choices of $\mathbf{P}$ are an all-ones matrix
to penalize all the elements of $\boldsymbol{\Sigma}$, and an all-ones
matrix with the diagonal elements set to zero to avoid penalizing
the diagonal elements of $\boldsymbol{\Sigma}$). The problem in (\ref{eq:5-1})
is challenging since the negative log-likelihood of the covariance
matrix is not convex. The authors of \cite{bien2011sparse} and \cite{wang2014coordinate}
estimated sparse covariance matrices by solving (\ref{eq:5-1}) using
a majorize-minimize method (SPCOV) and a coordinate descent approach,
respectively. However, in general, the $\ell_{1}$ norm penalty causes
shrinkage towards the origin leading to biased estimates. A better
way to impose sparsity is to use the $\ell_{0}$ norm penalty in the
MLE problem:
\begin{equation}
\underset{\boldsymbol{\Sigma}\succ0}{\min}\;\log\det\left(\boldsymbol{\Sigma}\right)+\textrm{Tr}\left(\boldsymbol{\Sigma}^{-1}\mathbf{S}\right)+\lambda\left\Vert \boldsymbol{\Sigma}\right\Vert _{0},\label{eq:6-1}
\end{equation}
where $\left\Vert \boldsymbol{\Sigma}\right\Vert _{0}$ denotes the
number of nonzero off-diagonal elements in $\boldsymbol{\Sigma}$.
However, solving (\ref{eq:6-1}) is even more challenging as both
the negative log-likelihood and the penalty terms are non-convex.
The authors of \cite{phan2017sparse} replaced the $\ell_{0}$ norm
in (\ref{eq:6-1}) with non-convex DC (difference of convex functions)
approximations and solved the resultant approximate problem using
DC programming techniques. Recently, the authors of \cite{xu2022proximal}
proposed an alternative method for solving problem (\ref{eq:6-1}).
Specifically, they minimized the negative log-likelihood subject to
the constraint that $\left\Vert \boldsymbol{\Sigma}\right\Vert _{0}\leq2k$,
where $2k$ denotes the desired number of non-zero off-diagonal elements
in the covariance matrix, using a distance-to-set penalty (Euclidean
distance from the target matrix to the constraint set) in place of
the sparsity inducing $\ell_{0}$ norm:
\begin{equation}
\underset{\boldsymbol{\Sigma}}{\min}\;\log\det\left(\boldsymbol{\Sigma}\right)+\textrm{Tr}\left(\boldsymbol{\Sigma}^{-1}\mathbf{S}\right)+\frac{\rho}{2}\textrm{dist}^{2}\left(\boldsymbol{\Sigma},\mathcal{C}\right).\label{eq:8}
\end{equation}
Here $\textrm{dist}^{2}\left(\boldsymbol{\Sigma},\mathcal{C}\right)$
is the minimum Euclidean distance from $\boldsymbol{\Sigma}$ to the
constraint set $\mathcal{C}=\left\{ \boldsymbol{\Sigma}|\boldsymbol{\Sigma}=\boldsymbol{\Sigma}^{T},\left\Vert \boldsymbol{\Sigma}\right\Vert _{0}\leq2k\right\} $.
Note that the constraint set consists of symmetric matrices with at
most $2k$ non-zero off-diagonal elements but does not impose positive-definiteness.
Moreover, the surrogate function used in \cite{xu2022proximal} is
constructed using local quadratic approximations and is not an actual
upper bound on the negative log-likelihood function. Consequently,
the authors of \cite{xu2022proximal} had to resort to back tracking
line search to ensure a monotonic decrease of the objective as well
as the positive definiteness of the iterates. 

Apart from imposing sparsity on the covariance matrix, a number of
papers impose sparsity on the inverse covariance matrix $\boldsymbol{\Phi}\triangleq\boldsymbol{\Sigma}^{-1}$.
Similar to the penalized formulation in (\ref{eq:5-1}), the problem
of sparse inverse covariance matrix estimation is:
\begin{equation}
\underset{\boldsymbol{\Phi}\succ0}{\min}\;-\log\det\left(\boldsymbol{\Phi}\right)+\textrm{Tr}\left(\boldsymbol{\Phi}\mathbf{S}\right)+\lambda\left\Vert \boldsymbol{\Phi}\right\Vert _{1}.\label{eq:8-1}
\end{equation}
This is a convex problem whose global minimizer can be obtained using
solvers like CVX \cite{cvx} or iterative techniques like the graphical
Lasso \cite{friedman2008sparse}. All the aforementioned methods \cite{bien2011sparse,wang2014coordinate,phan2017sparse,xu2022proximal}
including the penalized sparse inverse covariance matrix approach
in (\ref{eq:8-1}) require the appropriate selection of tuning parameters
($\lambda$ and $k$) which is usually done via multi-fold cross-validation
in which for every value of the parameter in its range the problem
must be solved many times and this can become intractable for large
dimensions. 

Regarding the maximum likelihood estimation in covariance graphical
models the published approaches include the iterative conditional
fitting algorithm in \cite{drton2003new,chaudhuri2007estimation},
finding gene regulatory networks in \cite{butte2000discovering} and
identification of multi-factor models in \cite{grzebyk2004identification}.

\subsection{Contributions}

In this paper, we consider the problem of sparse covariance matrix
estimation and propose two methods to solve it. In contrast to the
state-of-the-art techniques, the proposed methods comprise two steps.
In the first step, which is common to both methods, we infer the bi-directed
graphical model (the underlying covariance graph) using multiple hypothesis
testing, and in the second step we solve the MLE problem based on
the inferred graph using two different approaches. Specifically, we
determine the sparsity-pattern using the multiple hypothesis testing
method of false discovery rate (FDR) applied to $\mathbf{S}$. Then
in the first method, we solve the MLE problem using a block coordinate
descent (BCD) approach to estimate the non-zero values. In the second
method, we employ a proximal distance (PD) algorithm that uses a penalty
in the ML objective to minimize the distance between the estimated
graphical model and the target covariance matrix. A novelty of our
framework lies in the use of hypothesis testing for estimating the
sparsity of the target covariance matrix, which also involves choosing
the FDR-controlling parameter. As we explain in the next section,
making this choice is computationally much less complex than the cross-validation
method used in other methods. The main contributions of the paper
are summarized below:
\begin{enumerate}
\item We propose a two-step sparse covariance matrix estimation framework
that infers the sparsity pattern through multiple hypothesis testing
(FDR) to find the covariance graph and then solves the MLE problem
using two different methods (BCD and PD) based on the estimated covariance
graph. 
\item The proposed BCD method can solve problem (\ref{eq:6-1}) which is
solved via DC approximations in \cite{phan2017sparse}. In addition,
our PD method is a more accurate approach for solving (\ref{eq:6-1})
than the proximal distance algorithm in \cite{xu2022proximal}. 
\item We test the performance of the proposed approach on both simulated
data as well as real-world data (international migration data and
cell signaling data) and compare the results with those corresponding
to the sample covariance matrix and the state-of-the-art methods in
\cite{bien2011sparse} and \cite{xu2022proximal}.
\end{enumerate}
\emph{Notations}: Scalars, column vectors and matrices are denoted
by italics ($a$), bold small letters ($\mathbf{a}$) and bold capital
letters ($\mathbf{A}$), respectively. The $i^{\textrm{th}}$ element
of a vector $\mathbf{a}$ is denoted $a_{i}$, and the $i^{\textrm{th}}$
column and $\left(i,j\right)^{\textrm{th}}$element of matrix $\mathbf{A}$
is denoted $\mathbf{a}_{i}$ and $A_{ij}$. The notations $\left(\cdot\right)^{T}$,
$\left(\cdot\right)^{-1}$, $\det\left(\cdot\right)\slash\left|\cdot\right|$,
$\textrm{Tr}\left(\cdot\right)$ and $\left\Vert \cdot\right\Vert _{F}$
are used for the transpose, inverse, determinant, trace and Frobenius
norm of a matrix. $\mathbf{A}\succ0$ denotes a positive definite
matrix. $\hat{\mathbf{A}}$ denotes the estimated matrix $\mathbf{A}$,
and $\mathbf{A}^{t}$ denotes the matrix $\mathbf{A}$ at the end
of the $t^{\textrm{th}}$ iteration. $\mathbb{E}\left(\cdot\right)$
denotes expectation and $\mathbb{R}$ denotes the field of real numbers.
$\mathcal{G}\left(\mathcal{V},\mathcal{E},\mathbf{Z}\right)$ denotes
a graph where $\mathcal{V}$ is the set of vertices, $\mathcal{E}$
is the set of edges, and $\mathbf{Z}$ is the adjacency matrix. The
element-wise sparsity indicator function $\mathbb{I}\left(\mathbf{X}\right)$
is defined as:
\[
\mathbb{I}\left(X_{ij}\right)=\begin{cases}
0 & \textrm{ if }X_{ij}=0\;\textrm{must hold}\\
1 & \textrm{ otherwise }
\end{cases}
\]

\section{Proposed Methodology}

The proposed sparse covariance estimation approach is a two-stage
procedure. In the first stage, we use hypothesis testing to estimate
the sparsity-pattern of the target covariance matrix in the form of
a binary matrix $\mathbf{Z}$. Specifically we use the FDR method
of multiple hypothesis testing for sparsity pattern estimation. In
the second stage, we solve the MLE problem in (\ref{eq:3}) under
the inferred sparsity pattern using either a BCD method or a PD method.

We first present the FDR-based hypothesis testing for inferring the
covariance graph and then present the coordinate descent and proximal
distance algorithms to solve the MLE problem. Lastly, we discuss the
computational complexity, convergence and initialization of the proposed
algorithms.

\subsection{\label{subsec:Hypothesis-testing}First stage: Hypothesis testing}

We frame the problem of detecting whether an element $\Sigma_{ij}$
of $\boldsymbol{\Sigma}$ is zero or not as a hypothesis testing problem:
\begin{equation}
H_{ij}:\textrm{\ensuremath{\Sigma_{ij}}= 0},\;i,j=1,\ldots,p,\;i\neq j\label{eq:4}
\end{equation}
We will use Pearson's correlation coefficient \cite{dunn1986applied}
(denoted $\rho_{ij}$) to test the null hypothesis $H_{ij}$ :
\begin{equation}
\rho_{ij}=\frac{S_{ij}}{\sqrt{S_{ii}S_{jj}}}
\end{equation}
We compute the following test statistic:
\begin{equation}
T_{ij}\triangleq\rho_{ij}\sqrt{\frac{n-2}{1-\rho_{ij}^{2}}}\label{eq:6}
\end{equation}
which follows a Student's t-distribution with $n-2$ degrees of freedom
under the null hypothesis \cite{rahman}. In the proposed algorithms,
there are a total of $M=p(p-1)/2$ hypotheses to be tested (the upper
or lower triangular elements of $\boldsymbol{\Sigma}$). For ease
of notation, we change the indexing of the hypotheses from using two
indexes to a single index and henceforth the hypotheses are denoted
as $\left\{ H_{m}\right\} _{m=1}^{M}$. 

\subsubsection{False Discovery rate (FDR)}

An estimate of the sparsity pattern of $\boldsymbol{\Sigma}$ can
be obtained using any testing procedure that determines which of the
hypotheses in (\ref{eq:4}) should be accepted. It is in principle
possible to utilize even an individual testing process that evaluates
each hypothesis separately. However, because of the accumulation of
the false alarm probabilities of the individual tests, such a procedure
may have a high overall false alarm probability. A procedure that
can test all the hypotheses simultaneously and control the total false
alarm probability is therefore advised. To this end, we utilize the
FDR \cite{benjamini1995controlling,benjamini2001control}, which controls
the expected ratio of the number of incorrectly rejected hypotheses
($IR$) to the total number of rejected hypotheses ($R$):
\begin{equation}
\textrm{FDR}=\mathbb{E}\left(\frac{IR}{R}\right)
\end{equation}
The first step of the FDR is to rearrange the test statistics $\left\{ T_{m}\right\} $
in a descending order:
\begin{equation}
T_{[1]}\geq T_{[2]}\geq\cdots\geq T_{[M]},\label{eq:11}
\end{equation}
where $\left[\cdot\right]$ denotes the ordered indices. The significance
levels used by FDR are given by \cite{benjamini1995controlling}:
\begin{equation}
p_{m}=\alpha\frac{m}{M\eta_{M}},\label{eq:12}
\end{equation}
where alpha is a pre-specified value that controls the desired level
of FDR ($\textrm{FDR}\leq\alpha$) and $\eta_{m}$ is the harmonic
number defined as $\eta_{M}=\stackrel[i=1]{M}{\sum}\frac{1}{i}$ ($\approx\log M+0.577$
for large $M$). The quantiles $\left\{ q_{p_{m}}\right\} $ corresponding
to (\ref{eq:12}) are calculated using the Student's t-distribution
of the test statistics:
\begin{equation}
q_{p_{m}}:\textrm{prob}\left(T_{\left[m\right]}\geq q_{p_{m}}|H_{m}\right)=p_{m}
\end{equation}
 Let 
\begin{equation}
m_{\textrm{max}}=\textrm{max}\left[m:T_{m}\geq q_{p_{m}}\right].\label{eq:13}
\end{equation}
Then, we reject $H_{i}$ for $i=1,\ldots,m_{\textrm{max}}$ and accept
$H_{i}$ for $i=m_{\textrm{max}}+1,\ldots,m$, or reject no hypothesis
if no $m$ satisfies (\ref{eq:13}). We construct the matrix $\mathbf{Z}$,
which encodes the sparsity pattern, placing zeros at the indices corresponding
to the accepted hypotheses, and $1$ at the remaining indices.

\subsubsection{\label{subsec:Choosing-the-optimal}Choosing the value of $\alpha$
in FDR}

We employ the model selection rule of Extended Bayesian information
Criterion (EBIC) \cite{chen2008extended} to choose $\alpha$. As
a first step, we grid the interval $\left[0,1\right]$ using $K$
points (say, with a spacing of $0.005$) and construct a set $\boldsymbol{\alpha}\triangleq\left\{ \alpha_{k}\right\} _{k=1}^{K}$.
For each $k$, we then run FDR with $\alpha_{k}$ and obtain the sparse
pattern matrices $\left\{ \mathbf{Z}_{k}\right\} $. Typically many
values of $\alpha_{k}$'s give the same sparsity pattern. It is also
important to note that the sparsity patterns obtained using FDR for
increasing values of $\alpha$ are hierarchical, i.e., $\textrm{supp}\left(Z_{k-1}\right)\subseteq\textrm{supp}\left(Z_{k}\right)$
for $\alpha_{k-1}\leq\alpha_{k}$. Therefore, from the set $\boldsymbol{\alpha}$
we obtain a new set $\overline{\boldsymbol{\alpha}}$ with only those
values of $\alpha_{k}$ that yield different patterns $\mathbf{Z}_{k}$.
For each value of $\alpha_{k}$ in $\overline{\boldsymbol{\alpha}},$
we calculate the MLE $\hat{\boldsymbol{\Sigma}}_{k}$ of $\boldsymbol{\Sigma}$
(using the methods discussed in the next section) and the EBIC criterion:
\begin{equation}
\textrm{EBIC}_{k}=-2\mathcal{L}\left(\hat{\boldsymbol{\Sigma}}_{k}\right)+m_{k}\log\left(pn\right)+2m_{k}\log\left(M+p\right),\label{eq:13-1}
\end{equation}
where $\mathcal{L}\left(\cdot\right)$ is the log-likelihood function
of the multivariate Gaussian data as defined in (\ref{eq:1}) and
$\small m_{k}\log\left(pn\right)+2m_{k}\log\left(M+p\right)$ is the
EBIC penalty in which $m_{k}$ denotes the number of non-zero elements
in $\hat{\boldsymbol{\Sigma}}_{k}$. 

In Section III, we will numerically illustrate the use of FDR for
choosing $\alpha$. We note here that the EBIC based choice of $\alpha$
is not as cumbersome from a computational view point as using cross-validation
for choosing the penalty parameter ($\lambda$) in \cite{bien2011sparse}
or the sparsity level ($k$) in \cite{xu2022proximal}. Firstly, the
search for $\alpha$ is limited to a sub-interval of $[0,1]$ (usually
$[0,0.1]$) which is not the case in the aforementioned methods where
$\lambda>0$ in \cite{bien2011sparse} and $0\leq k\leq$$M$ in \cite{xu2022proximal}.
Secondly, unlike the methods in the cited papers, in the proposed
approach we need to solve the MLE problem only for a small number
of values of $\alpha_{k}$ (typically less than $10$, see Fig \ref{fig:ebic}
in Section \ref{sec:Numerical-Simulations}). 

\subsection{Second stage: Solving the MLE problem}

Now that we have determined a sparsity pattern of the underlying covariance
matrix, we move on to estimate the non-zero $\left\{ \Sigma_{ij}\right\} $
by solving the constrained problem in (\ref{eq:3}) using two different
methods. 

\subsubsection{Block Coordinate descent (BCD) method}

We first present a brief overview of the general BCD method, and then
discuss the proposed BCD algorithm in detail.

\paragraph{Block Coordinate descent}

BCD is an iterative approach which can be used to solve large dimensional
optimization problems of the form:
\begin{equation}
\begin{aligned}\underset{\mathbf{x}_{1},\mathbf{x}_{2},\ldots,\mathbf{x}_{n}}{\min}\; & f\left(\mathbf{x}_{1},\mathbf{x}_{2},\ldots,\mathbf{x}_{n}\right)\\
\mathrm{s.t.\;} & \mathbf{x}_{1},\mathbf{x}_{2},\ldots,\mathbf{x}_{n}\in\mathcal{X}
\end{aligned}
\end{equation}
At each iteration of BCD, the problem is solved with respect to one
block of the optimization variable, with the remaining blocks fixed
at their values from the previous iteration. The order in which the
blocks are updated can be either random or deterministic (i.e., according
to some coordinate selection rule). One of the simplest deterministic
way is to cyclically iterate through the variables and update them,
i.e., at the $\left(t+1\right)^{\textrm{th}}$ iteration:
\begin{equation}
\begin{aligned}\mathbf{x}_{i}^{t+1} & =\textrm{arg}\;\underset{\mathbf{x}_{i}\in\mathcal{X}}{\min}\;f\left(\mathbf{x}_{1}^{t},\mathbf{x}_{2}^{t},\ldots,\mathbf{x}_{i-1}^{t},\mathbf{x}_{i},\mathbf{x}_{i+1}^{t},\ldots,\mathbf{x}_{n}^{t}\right),\\
 & \,\,\,\,\,\,\,\,\,\,\,\,\,\,\,\,\,\,\,\,\,\,\,\,\,\,\,\,\,\,\,\,\,\,\,\,\,\,\,\,\,\,\,\,\,\,\,\,\,\,\,\,\,\,\,\,\,\,\,\,\,\,\,\,\,i=(t+1)\textrm{ mod }n
\end{aligned}
\end{equation}
\begin{align}
\mathbf{x}_{j}^{t+1} & =\mathbf{x}_{j}^{t},\;\forall j\neq i
\end{align}
The BCD method is useful in large-scale problems where a batch update
(updating all the variables together) would be too complex and computationally
expensive.

\paragraph{The Proposed BCD method for sparse covariance matrix estimation}

Let us first restate the problem (\ref{eq:3}):
\begin{equation}
\begin{aligned}\underset{\boldsymbol{\Sigma}\succ0}{\min}\; & \log\det\left(\boldsymbol{\Sigma}\right)+\textrm{Tr}\left(\boldsymbol{\Sigma}^{-1}\mathbf{S}\right)\\
\mathrm{s.t.\;} & \boldsymbol{\Sigma}\in\mathcal{C},
\end{aligned}
\label{eq:19-2}
\end{equation}
where $\mathcal{C}\triangleq\left\{ \boldsymbol{\Sigma}|\mathbb{I}\left(\boldsymbol{\Sigma}\right)=\mathbf{Z}\right\} $
is the constraint set. Problem (\ref{eq:19-2}) is solved using a
block coordinate descent based method.

In the proposed method, we update a $2\times2$ block of the optimization
variable $\boldsymbol{\Sigma}$ at each iteration. The block to be
updated is selected cyclically as follows: Let $\mathcal{V}=\left\{ 1,2,\ldots p\right\} $
be the set of vertices (variables), and $\mathcal{A}=\{u,v\}\subset\mathcal{V}$
be the subset that describes the block to be updated. Then, the optimization
variable $\boldsymbol{\Sigma}$ at the $\left(t+1\right)^{\textrm{th}}$
iteration can be written as follows:
\begin{equation}
\boldsymbol{\Sigma}=\begin{bmatrix}\boldsymbol{\Sigma}_{\mathcal{AA}} & \boldsymbol{\Sigma}_{\mathcal{AB}}^{t}\\
\boldsymbol{\Sigma}_{\mathcal{BA}}^{t} & \boldsymbol{\Sigma}_{\mathcal{BB}}^{t}
\end{bmatrix}\label{eq:22-1}
\end{equation}
where $\mathcal{AA}=\mathcal{A}\times\mathcal{A}$ gives the indices
of the block to be updated at $\left(t+1\right)^{\textrm{th}}$ iteration
and $\mathcal{AB}=\mathcal{A}\times\mathcal{B}$, $\mathcal{BA}=\mathcal{B}\times\mathcal{A}$
and $\mathcal{BB}=\mathcal{B}\times\mathcal{B}$ give the indices
of the updated blocks from the previous $t^{\textrm{th}}$ iteration.
It is worth mentioning that the block $\boldsymbol{\Sigma}_{\mathcal{AA}}$
to be updated can lie anywhere in the matrix $\boldsymbol{\Sigma}$.
However for the ease of derivation and notation, we move it to the
top-left corner using a permutation matrix $\mathbf{U}$. The matrix
$\mathbf{U}$ is constructed by interchanging the $u^{\textrm{th}}$
and $v^{\textrm{th}}$ column of an identity matrix with its first
and second columns, respectively. Then, the optimization variable
$\boldsymbol{\Sigma}$ as well as the matrices $\mathbf{S}$ and $\mathbf{Z}$
are pre-multiplied with $\mathbf{U}^{T}$ and post-multiplied with
$\mathbf{U}$ for rearrangement of the elements. 

Thus, the problem to be solved at the $\left(t+1\right)^{\textrm{th}}$
iteration can be written as:

\begin{equation}
\begin{aligned}\underset{\boldsymbol{\Sigma}_{AA}\succ0}{\min}\; & \textrm{Tr}\left(\begin{bmatrix}\boldsymbol{\Sigma}_{\mathcal{AA}} & \boldsymbol{\Sigma}_{\mathcal{AB}}^{t}\\
\boldsymbol{\Sigma}_{\mathcal{BA}}^{t} & \boldsymbol{\Sigma}_{\mathcal{BB}}^{t}
\end{bmatrix}^{-1}\begin{bmatrix}\mathbf{S}_{\mathcal{AA}} & \mathbf{S}_{\mathcal{AB}}\\
\mathbf{S}_{\mathcal{BA}} & \mathbf{S}_{\mathcal{BB}}
\end{bmatrix}\right)\\
 & +\log\begin{vmatrix}\boldsymbol{\Sigma}_{\mathcal{AA}} & \boldsymbol{\Sigma}_{\mathcal{AB}}^{t}\\
\boldsymbol{\Sigma}_{\mathcal{BA}}^{t} & \boldsymbol{\Sigma}_{\mathcal{BB}}^{t}
\end{vmatrix}\\
\mathrm{s.t.\;} & \mathbb{I}\left(\boldsymbol{\Sigma}_{\mathcal{AA}}\right)=\mathbf{Z}_{\mathcal{AA}},
\end{aligned}
\label{eq:15}
\end{equation}
where $\mathbf{Z}_{\mathcal{AA}}\in[0,1]^{2\times2}$ is the upper
left block of the matrix $\mathbf{Z}$. Next, we express $\left|\boldsymbol{\Sigma}\right|$
in terms of the variable $\boldsymbol{\Sigma}_{AA}$ as shown below:
\begin{equation}
\begin{aligned}\left|\boldsymbol{\Sigma}\right| & =\begin{vmatrix}\mathbf{I} & \boldsymbol{\Sigma}_{\mathcal{AB}}^{t}\\
\mathbf{0} & \boldsymbol{\Sigma}_{\mathcal{BB}}^{t}
\end{vmatrix}\begin{vmatrix}\boldsymbol{\Sigma}_{\mathcal{AA}}-\boldsymbol{\Sigma}_{\mathcal{AB}}^{t}\left(\boldsymbol{\Sigma}_{\mathcal{BB}}^{t}\right)^{-1}\boldsymbol{\Sigma}_{\mathcal{BA}}^{t} & \mathbf{0}\\
\left(\boldsymbol{\Sigma}_{\mathcal{BB}}^{t}\right)^{-1}\boldsymbol{\Sigma}_{\mathcal{BA}}^{t} & \mathbf{I}
\end{vmatrix}\\
 & =\left|\boldsymbol{\Sigma}_{\mathcal{BB}}^{t}\right|\left|\boldsymbol{\Sigma}_{\mathcal{AA}}-\boldsymbol{\Sigma}_{\mathcal{AB}}^{t}\left(\boldsymbol{\Sigma}_{\mathcal{BB}}^{t}\right)^{-1}\boldsymbol{\Sigma}_{\mathcal{BA}}^{t}\right|\\
 & =\left|\boldsymbol{\Sigma}_{\mathcal{BB}}^{t}\right|\left|\boldsymbol{\Sigma}_{\mathcal{AA}}-\boldsymbol{\Psi}\right|\\
 & =\left|\boldsymbol{\Sigma}_{\mathcal{BB}}^{t}\right|\left|\bar{\boldsymbol{\Sigma}}\right|
\end{aligned}
\label{eq:22}
\end{equation}
where $\boldsymbol{\Psi}\triangleq\boldsymbol{\Sigma}_{\mathcal{AB}}^{t}\left(\boldsymbol{\Sigma}_{\mathcal{BB}}^{t}\right)^{-1}\boldsymbol{\Sigma}_{\mathcal{BA}}^{t}$
and $\bar{\boldsymbol{\Sigma}}\triangleq\boldsymbol{\Sigma}_{\mathcal{AA}}-\boldsymbol{\Psi}$.
Similarly, 
\begin{equation}
\begin{aligned}\boldsymbol{\Sigma}^{-1} & =\begin{bmatrix}\bar{\boldsymbol{\Sigma}} & \mathbf{0}\\
\left(\boldsymbol{\Sigma}_{\mathcal{BB}}^{t}\right)^{-1}\boldsymbol{\Sigma}_{\mathcal{BA}}^{t} & \mathbf{I}
\end{bmatrix}^{-1}\begin{bmatrix}\mathbf{I} & \boldsymbol{\Sigma}_{\mathcal{AB}}^{t}\\
\mathbf{0} & \boldsymbol{\Sigma}_{\mathcal{BB}}^{t}
\end{bmatrix}^{-1}\\
 & =\begin{bmatrix}\bar{\boldsymbol{\Sigma}}^{-1} & \mathbf{0}\\
-\left(\boldsymbol{\Sigma}_{\mathcal{BB}}^{t}\right)^{-1}\boldsymbol{\Sigma}_{\mathcal{BA}}^{t}\bar{\boldsymbol{\Sigma}}^{-1} & \mathbf{I}
\end{bmatrix}\begin{bmatrix}\mathbf{I} & -\boldsymbol{\Sigma}_{\mathcal{AB}}^{t}\left(\boldsymbol{\Sigma}_{\mathcal{BB}}^{t}\right)^{-1}\\
\mathbf{0} & \left(\boldsymbol{\Sigma}_{\mathcal{BB}}^{t}\right)^{-1}
\end{bmatrix}\\
 & =\small\begin{bmatrix}\bar{\boldsymbol{\Sigma}}^{-1} & -\bar{\boldsymbol{\Sigma}}^{-1}\boldsymbol{\Sigma}_{\mathcal{AB}}^{t}\left(\boldsymbol{\Sigma}_{\mathcal{BB}}^{t}\right)^{-1}\\
-\left(\boldsymbol{\Sigma}_{\mathcal{BB}}^{t}\right)^{-1}\boldsymbol{\Sigma}_{\mathcal{BA}}^{t}\bar{\boldsymbol{\Sigma}}^{-1} & \left(\left(\boldsymbol{\Sigma}_{\mathcal{BB}}^{t}\right)^{-1}\boldsymbol{\Sigma}_{\mathcal{BA}}^{t}\bar{\boldsymbol{\Sigma}}^{-1}\boldsymbol{\Sigma}_{\mathcal{AB}}^{t}\right.\\
 & \left.\left(\boldsymbol{\Sigma}_{\mathcal{BB}}^{t}\right)^{-1}+\left(\boldsymbol{\Sigma}_{\mathcal{BB}}^{t}\right)^{-1}\right)
\end{bmatrix}\\
 & =\begin{bmatrix}\bar{\boldsymbol{\Sigma}}^{-1} & -\bar{\boldsymbol{\Sigma}}^{-1}\boldsymbol{\Phi}^{T}\\
-\boldsymbol{\Phi}\bar{\boldsymbol{\Sigma}}^{-1} & \boldsymbol{\Phi}\bar{\boldsymbol{\Sigma}}^{-1}\boldsymbol{\Phi}^{T}+\left(\boldsymbol{\Sigma}_{\mathcal{BB}}^{t}\right)^{-1}
\end{bmatrix},
\end{aligned}
\label{23}
\end{equation}
where $\boldsymbol{\Phi}\triangleq\left(\boldsymbol{\Sigma}_{\mathcal{BB}}^{t}\right)^{-1}\boldsymbol{\Sigma}_{\mathcal{BA}}^{t}$.
Using (\ref{eq:22}) and (\ref{23}) in (\ref{eq:15}), and ignoring
the constants, we get:
\begin{equation}
\begin{aligned}\underset{\bar{\boldsymbol{\Sigma}}\succ0}{\min}\; & \textrm{Tr}\left(\begin{bmatrix}\bar{\boldsymbol{\Sigma}}^{-1} & -\bar{\boldsymbol{\Sigma}}^{-1}\boldsymbol{\Phi}^{T}\\
-\boldsymbol{\Phi}\bar{\boldsymbol{\Sigma}}^{-1} & \boldsymbol{\Phi}\bar{\boldsymbol{\Sigma}}^{-1}\boldsymbol{\Phi}^{T}+\left(\boldsymbol{\Sigma}_{\mathcal{BB}}^{t}\right)^{-1}
\end{bmatrix}\mathbf{S}\right)+\log\left|\bar{\boldsymbol{\Sigma}}\right|\\
\mathrm{s.t.\;} & \mathbb{I}\left(\bar{\boldsymbol{\Sigma}}+\boldsymbol{\Psi}\right)=\mathbf{Z}_{\mathcal{AA}}
\end{aligned}
\label{eq:16}
\end{equation}
which can be compactly written as:
\begin{equation}
\begin{aligned}\underset{\bar{\boldsymbol{\Sigma}}\succ0}{{\rm min}}\: & g\left(\bar{\boldsymbol{\Sigma}}\right)\triangleq\mathrm{Tr}\left(\boldsymbol{\Theta}\bar{\boldsymbol{\Sigma}}^{-1}\right)+\mathrm{log}\left|\bar{\boldsymbol{\Sigma}}\right|\\
\mathrm{s.t.\;} & \mathbb{I}\left(\bar{\boldsymbol{\Sigma}}+\boldsymbol{\Psi}\right)=\mathbf{Z}_{\mathcal{AA}}
\end{aligned}
\label{eq:17}
\end{equation}
where $\boldsymbol{\Theta}\triangleq\mathbf{S}_{\mathcal{AA}}-\mathbf{S}_{\mathcal{AB}}\boldsymbol{\Phi}-\boldsymbol{\Phi}^{T}\mathbf{S}_{\mathcal{BA}}+\boldsymbol{\Phi}^{T}\mathbf{S}_{\mathcal{BB}}\boldsymbol{\Phi}$.
Since $\boldsymbol{\Theta}$ is positive-definite, $g\left(\bar{\boldsymbol{\Sigma}}\right)$
is bounded from below. Moreover $g\left(\bar{\boldsymbol{\Sigma}}\right)$
is convex in $\bar{\boldsymbol{\Sigma}}^{-1}$ and therefore has a
unique minimum $\hat{\bar{\boldsymbol{\Sigma}}}=\boldsymbol{\Theta}$
(and $\hat{\boldsymbol{\Sigma}}_{\mathcal{AA}}=\boldsymbol{\Theta}+\boldsymbol{\Psi}$),
which will also be the solution of the constrained problem (\ref{eq:17})
if the constraint $\mathbb{I}\left(\boldsymbol{\Theta}+\boldsymbol{\Psi}\right)=\mathbf{Z}_{\mathcal{AA}}$
is satisfied. To solve the constrained problem (\ref{eq:17}) we proceed
as follows.

If $\left(\mathbf{Z}_{\mathcal{AA}}\right)_{12}=1$, then the element
$\bar{\boldsymbol{\Sigma}}_{12}+\boldsymbol{\Psi}_{12}$ may or may
not be zero, i.e., there is no restriction on the solution and the
block $\boldsymbol{\mathbf{\Sigma}}_{\mathcal{AA}}$ is updated as
$\boldsymbol{\Sigma}_{\mathcal{AA}}^{t+1}=\boldsymbol{\Theta}+\boldsymbol{\Psi}$.
However, if $\left(\mathbf{Z}_{\mathcal{AA}}\right)_{12}=0$, then
to satisfy the constraint the optimal solution is (note that $\bar{\boldsymbol{\Sigma}}$
and $\boldsymbol{\Psi}$ are $2\times2$ matrices):
\begin{equation}
\hat{\bar{\Sigma}}_{12}=-\Psi_{12}\label{eq:18}
\end{equation}
To find the optimal values of $\bar{\Sigma}_{11}$ and $\bar{\Sigma}_{22}$
in this case, we rewrite equation (\ref{eq:17}) using $\bar{\Sigma}_{12}=-\Psi_{12}$
to get:
\begin{equation}
\begin{aligned}\underset{\bar{\boldsymbol{\Sigma}}\succ0}{{\rm min}}\: & \mathrm{Tr}\left(\begin{bmatrix}\Theta_{11} & \Theta_{12}\\
\Theta_{12} & \Theta_{22}
\end{bmatrix}\begin{bmatrix}\bar{\Sigma}_{11} & -\Psi_{12}\\
-\Psi_{12} & \bar{\Sigma}_{22}
\end{bmatrix}^{-1}\right)\\
 & +\mathrm{log}\begin{vmatrix}\bar{\Sigma}_{11} & -\Psi_{12}\\
-\Psi_{12} & \bar{\Sigma}_{22}
\end{vmatrix}
\end{aligned}
\label{eq:19}
\end{equation}
or equivalently
\begin{equation}
\begin{aligned}\underset{\bar{\boldsymbol{\Sigma}}\succ0}{{\rm min}}\: & \frac{1}{\bar{\Sigma}_{11}\bar{\Sigma}_{22}-\Psi_{12}^{2}}\left(\Theta_{11}\bar{\Sigma}_{22}+2\Theta_{12}\Psi_{12}+\Theta_{22}\bar{\Sigma}_{11}\right)\\
 & +\mathrm{log}\left(\bar{\Sigma}_{11}\bar{\Sigma}_{22}-\Psi_{12}^{2}\right)
\end{aligned}
\label{eq:19-1}
\end{equation}
Note that the constraint $\bar{\boldsymbol{\Sigma}}\succ0$ in (\ref{eq:19})
is a convex set and the cost function in (\ref{eq:19}) is bounded
from below over the set. Additionally, on the boundary of this set
the objective function in (\ref{eq:19}) tends to $+\infty$. Therefore,
the problem in (\ref{eq:19}) has a bounded global minimum. 

Differentiating (\ref{eq:19-1}) with respect to $\bar{\Sigma}_{11}$
gives:
\begin{equation}
\Theta_{22}+\bar{\Sigma}_{22}=\frac{\left(\Theta_{11}\bar{\Sigma}_{22}+2\Theta_{12}\Psi_{12}+\Theta_{22}\bar{\Sigma}_{11}\right)\bar{\Sigma}_{22}}{\left(\bar{\Sigma}_{11}\bar{\Sigma}_{22}-\Psi_{12}^{2}\right)}\label{eq:29}
\end{equation}
Similarly, differentiating (\ref{eq:19-1}) with respect to $\bar{\Sigma}_{22}$
gives:
\begin{equation}
\Theta_{11}+\bar{\Sigma}_{11}=\frac{\left(\Theta_{11}\bar{\Sigma}_{22}+2\Theta_{12}\Psi_{12}+\Theta_{22}\bar{\Sigma}_{11}\right)\bar{\Sigma}_{11}}{\left(\bar{\Sigma}_{11}\bar{\Sigma}_{22}-\Psi_{12}^{2}\right)}\label{eq:30}
\end{equation}
Dividing (\ref{eq:29}) by (\ref{eq:30}) yields:
\begin{equation}
\bar{\Sigma}_{22}=\frac{\Theta_{22}}{\Theta_{11}}\bar{\Sigma}_{11}\label{eq:31}
\end{equation}
Substituting for $\bar{\Sigma}_{22}$ in (\ref{eq:29}) we have that:
\begin{equation}
\Theta_{22}\bar{\Sigma}_{11}^{3}-\Theta_{22}\Theta_{11}\bar{\Sigma}_{11}^{2}-\Theta_{11}(\Psi_{12}^{2}+2\Theta_{12}\Psi_{12})\bar{\Sigma}_{11}-\Psi_{12}^{2}\Theta_{11}^{2}=0\label{eq:32}
\end{equation}
which is a cubic equation in $\bar{\Sigma}_{11}$. Note that the first
coefficient ($\Theta_{22}$) is positive, and the second and the fourth
coefficients ($-\Theta_{22}\Theta_{11}$ and $-\Psi_{12}^{2}\Theta_{11}^{2}$)
are negative (since $\Theta_{11}$ and $\Theta_{22}$ are positive
quantities), therefore, depending on the sign of the third coefficient
$-\Theta_{11}(\Psi_{12}^{2}+2\Theta_{12}\Psi_{12})$, (\ref{eq:32})
can have either one or three positive solutions. If $\Psi_{12}^{2}+2\Theta_{12}\Psi_{12}$
is positive, then by Descartes' rule of signs, there is exactly one
positive solution of (\ref{eq:32}) which gives the optimal value
$\bar{\Sigma}_{11}.$ However, if $\Psi_{12}^{2}+2\Theta_{12}\Psi_{12}$
is negative, then there might be three positive solutions of which
the one that yields the lowest value of the objective in (\ref{eq:17})
gives the optimal value of $\bar{\Sigma}_{11}.$ Thus, in the present
case, the block $\mathbf{\Sigma}_{\mathcal{AA}}$ is updated as: 
\begin{equation}
\mathbf{\boldsymbol{\Sigma}}_{\mathcal{AA}}^{t+1}=\begin{bmatrix}\hat{\bar{\Sigma}}_{11}+\Psi_{11} & 0\\
0 & \hat{\bar{\Sigma}}_{22}+\Psi_{22}
\end{bmatrix}\label{eq:33}
\end{equation}
The stationary point of (\ref{eq:19}) thus obtained via the aforementioned
method is also the global minimizer. This proves that the covariance
matrix $\bar{\boldsymbol{\Sigma}}^{t+1}$ obtained at the end of each
iteration is positive definite. The matrix $\mathbf{\boldsymbol{\Sigma}}_{\mathcal{AA}}^{t+1}$
at each iteration is also positive definite because when $\left(\mathbf{Z}_{\mathcal{AA}}\right)_{12}=1$,
$\mathbf{\boldsymbol{\Sigma}}_{\mathcal{AA}}^{t+1}$ is equal to the
sum of two positive definite matrices $\boldsymbol{\Theta}$ and $\boldsymbol{\Psi}$,
and when $\left(\mathbf{Z}_{\mathcal{AA}}\right)_{12}=0$, $\mathbf{\boldsymbol{\Sigma}}_{\mathcal{AA}}^{t+1}$
is a diagonal matrix with positive elements $\hat{\bar{\Sigma}}_{11}+\Psi_{11}$
and $\hat{\bar{\Sigma}}_{22}+\Psi_{22}$ on the diagonal. Since both
$\mathbf{\boldsymbol{\Sigma}}_{\mathcal{AA}}^{t+1}$ and $\bar{\boldsymbol{\Sigma}}^{t+1}$
(which is the Schur complement of $\mathbf{\boldsymbol{\Sigma}}_{\mathcal{AA}}^{t+1}$)
are positive definite, the matrix $\mathbf{\boldsymbol{\Sigma}}^{t+1}$
as defined in (\ref{eq:22-1}) is also positive definite. We would
also like to note that when $\left(\mathbf{Z}_{\mathcal{AA}}\right)_{12}=1$,
the algorithm may introduce zeros in the estimated covariance matrix
in addition to those in the initial sparsity pattern constraint.

\begin{algorithm}[tbh]
\caption{\label{alg:Pseudo-code-CODE}The BCD algorithm}

\textbf{Input}: $\mathbf{S}$, $\alpha$ , $\epsilon$

\textbf{Initialize}: $t=0$, $\mathbf{\boldsymbol{\Sigma}}^{0}$ 
\begin{itemize}
\item Determine the sparsity pattern matrix $\mathbf{Z}$ using FDR as discussed
in Section \ref{subsec:Hypothesis-testing}.
\end{itemize}
\textbf{Iterate}

for $u=1:p-1$
\begin{description}
\item [{for}] $v=u+1:p$
\begin{itemize}
\item Compute $\mathbf{U}$ and rearrange the matrices $\boldsymbol{\Sigma}$,
$\mathbf{Z}$ and $\mathbf{S}$ using the permutation matrix $\mathbf{U}$.
\item Compute $\boldsymbol{\Theta}$ and $\boldsymbol{\Psi}$.
\item 

\textbf{If} $\left(Z_{\mathcal{AA}}\right)_{12}=1$, then:
\begin{itemize}
\item $\mathbf{\boldsymbol{\Sigma}}_{\mathcal{AA}}^{t+1}=\boldsymbol{\Theta}+\boldsymbol{\Psi}$
\end{itemize}
\textbf{Else}
\begin{itemize}
\item $\hat{\bar{\Sigma}}_{12}=-\Psi_{12}$
\item Solve (\ref{eq:32}) and (\ref{eq:31}) to obtain $\hat{\bar{\Sigma}}_{11}$
and $\hat{\bar{\Sigma}}_{22}$.
\item $\mathbf{\boldsymbol{\Sigma}}_{\mathcal{AA}}^{t+1}=\hat{\bar{\boldsymbol{\Sigma}}}+\boldsymbol{\Psi}$
\end{itemize}
\item $\boldsymbol{\Sigma}^{t+1}=\mathbf{U}\begin{bmatrix}\boldsymbol{\Sigma}_{\mathcal{AA}}^{t+1} & \boldsymbol{\Sigma}_{\mathcal{AB}}^{t}\\
\boldsymbol{\Sigma}_{\mathcal{BA}}^{t} & \boldsymbol{\Sigma}_{\mathcal{BB}}^{t}
\end{bmatrix}\mathbf{U}^{T}$
\end{itemize}
\item [{end}]~
\end{description}
end
\begin{itemize}
\item $t=t+1$
\end{itemize}
\textbf{Stop} and return $\boldsymbol{\Sigma}^{t+1}$ if $\left\Vert \boldsymbol{\Sigma}^{t+1}-\boldsymbol{\Sigma}^{t}\right\Vert _{F}\slash\left\Vert \boldsymbol{\Sigma}^{t}\right\Vert _{F}<\epsilon$

\textbf{Output}: $\hat{\boldsymbol{\Sigma}}=\boldsymbol{\Sigma}^{t+1}$
at convergence.
\end{algorithm}
Algorithm \ref{alg:Pseudo-code-CODE} summarizes the pseudo-code of
the BCD algorithm.
\begin{rem}
It is worth mentioning that the proposed method BCD can be used to
solve the $\ell_{0}$ norm regularized problem (\ref{eq:6-1}) after
a slight modification. The $2\times2$ problem in (\ref{eq:17}) can
be written with the $\ell_{0}$ norm penalty as: 
\begin{equation}
\begin{aligned}\underset{\bar{\boldsymbol{\Sigma}}\succ0}{{\rm min}}\: & \mathrm{Tr}\left(\boldsymbol{\Theta}\bar{\boldsymbol{\Sigma}}^{-1}\right)+\mathrm{log}\left|\bar{\boldsymbol{\Sigma}}\right|+\lambda\left\Vert \bar{\boldsymbol{\Sigma}}+\boldsymbol{\Psi}\right\Vert _{0}\end{aligned}
,\label{eq:33-1}
\end{equation}
for which two possible solutions exists: either the off-diagonal elements
of $\boldsymbol{\Sigma}_{\mathcal{AA}}=\bar{\boldsymbol{\Sigma}}+\boldsymbol{\Psi}$
are zero and thus $\bar{\Sigma}_{12}^{*}=\bar{\Sigma}_{21}^{*}=-\Psi_{12}$
and $\bar{\Sigma}_{11}^{*}$, $\bar{\Sigma}_{22}^{*}$ can be obtained
from (\ref{eq:31}) and (\ref{eq:32}), or the off-diagonal elements
of $\boldsymbol{\Sigma}_{\mathcal{AA}}$ are non-zero in which case
(\ref{eq:33-1}) has the closed-form solution $\bar{\boldsymbol{\Sigma}}^{*}=\boldsymbol{\Theta}$.
The solution that yields the lower value of the objective function
in (\ref{eq:2}) is chosen as the optimal solution. 
\end{rem}

\subsubsection{Proximal distance approach\label{subsec:Proximal-distance-algorithm}}

In this sub-section we first introduce the general proximal distance
algorithm, and then discuss the proposed PD method in detail. 

\paragraph{Proximal distance algorithm}

This type of algorithm is a combination of the classical Courant's
penalty method of constrained optimization and the principle of majorization-minimization
(MM). Consider a constrained optimization problem of the form:
\begin{equation}
\underset{\mathbf{x}\in\mathcal{C}}{\textrm{min}}\;f\left(\mathbf{x}\right),\label{eq:38}
\end{equation}
where $\mathcal{C}$ is the constraint set. Using the mentioned penalty
method the constrained problem (\ref{eq:38}) can be written as an
unconstrained problem as follows:
\begin{equation}
\underset{\mathbf{x}}{\textrm{min}}\;f\left(\mathbf{x}\right)+\rho q\left(\mathbf{x}\right)\label{eq:39}
\end{equation}
where the penalty $q\left(\mathbf{x}\right)$ is nonnegative and equal
to zero when $\mathbf{x}\in\mathcal{C}$. For $q\left(\mathbf{x}\right)\triangleq\frac{1}{2}\textrm{dist}^{2}\left(\mathbf{x},\mathcal{C}\right)$,
(\ref{eq:39}) can be written as: 
\begin{equation}
\underset{\mathbf{x}}{\textrm{min}}\;f\left(\mathbf{x}\right)+\frac{\rho}{2}\textrm{dist}^{2}\left(\mathbf{x},\mathcal{C}\right).\label{eq:40}
\end{equation}
Using the Euclidean distance in (\ref{eq:40}) we get:
\begin{equation}
\underset{\mathbf{x}}{\textrm{min}}\;f\left(\mathbf{x}\right)+\frac{\rho}{2}\left\Vert \mathbf{x}-P_{\mathcal{C}}\left(\mathbf{x}\right)\right\Vert ^{2},\label{eq:41}
\end{equation}
where $P_{\mathcal{C}}\left(\mathbf{x}\right)$ is the orthogonal
projection of $\mathbf{x}$ on $\mathcal{C}$ that minimizes the norm.
In the next step, the quadratic penalty is majorized using MM. Before
describing that step, we review the MM principle.

Let $h\left(\mathbf{x}\right)$ be the function to be minimized. In
the first step of MM, for a given $\mathbf{x}^{t}$ a surrogate $g\left(\mathbf{x}|\mathbf{x}^{t}\right)$
that has the following properties is constructed:
\begin{align}
g\left(\mathbf{x}|\mathbf{x}^{t}\right) & \geq h\left(\mathbf{x}\right)\;\textrm{and }\label{eq:42}\\
g\left(\mathbf{x}^{t}|\mathbf{x}^{t}\right) & =h\left(\mathbf{x}^{t}\right)\label{eq:43-1}
\end{align}
In the second step, the surrogate is minimized to obtain the next
update:
\begin{equation}
\mathbf{x}^{t+1}=\underset{\mathbf{x}}{\textrm{argmin }}g\left(\mathbf{x}|\mathbf{x}^{t}\right)\label{eq:44}
\end{equation}
From (\ref{eq:42}), (\ref{eq:43-1}), and (\ref{eq:44}), one can
verify that: 
\begin{equation}
h\left(\mathbf{x}^{t+1}\right)\leq g\left(\mathbf{x}^{t+1}|\mathbf{x}^{t}\right)\leq g\left(\mathbf{x}^{t}|\mathbf{x}^{t}\right)=h\left(\mathbf{x}^{t}\right).
\end{equation}
Therefore an MM algorithm monotonically decreases the objective function.
The key ingredient of such an algorithm is the construction of the
surrogate, which should closely follow the objective function and
should be simple to minimize. For details on the construction of surrogate
functions, we refer to \cite{sun2016majorization}.

Using the MM principle, a surrogate for the problem (\ref{eq:41})
is constructed majorizing the distance term:
\begin{equation}
g\left(\mathbf{x}|\mathbf{x}^{t}\right)\triangleq f\left(\mathbf{x}\right)+\frac{\rho}{2}\left\Vert \mathbf{x}-P_{\mathcal{C}}\left(\mathbf{x}^{t}\right)\right\Vert ^{2},\label{eq:44-1}
\end{equation}
For more details on proximal distance algorithms, the interested readers
can consult \cite{keys2019proximal}.

\paragraph{The Proposed PD method for sparse covariance matrix estimation}

The problem of sparse covariance matrix estimation with distance penalty
can be written as:
\begin{equation}
\underset{\boldsymbol{\Sigma}\succ0}{\textrm{min}}\log\left|\boldsymbol{\Sigma}\right|+\textrm{Tr}\left(\mathbf{\boldsymbol{\Sigma}}^{-1}\mathbf{S}\right)+\frac{\rho}{2}\textrm{dist}^{2}\left(\boldsymbol{\Sigma},\mathcal{C}\right)
\end{equation}
where $\textrm{dist}^{2}\left(\mathbf{\boldsymbol{\Sigma}},\mathcal{C}\right)$
is the squared distance from $\mathbf{\boldsymbol{\Sigma}}$ to $\mathcal{C}$
and $\rho$ is the penalty parameter. Writing the distance penalty
term using the Frobenius norm, we get:
\begin{equation}
\begin{aligned}\underset{\boldsymbol{\Sigma}\succ0}{\textrm{min}} & \log\left|\boldsymbol{\Sigma}\right|+\textrm{Tr}\left(\boldsymbol{\Sigma}^{-1}\mathbf{S}\right)+\frac{\rho}{2}\left\Vert \boldsymbol{\Sigma}-P_{\mathcal{C}}\left(\boldsymbol{\Sigma}\right)\right\Vert _{F}^{2}\end{aligned}
,\label{eq:2-1}
\end{equation}
 Problem (\ref{eq:2-1}) can also be written as:
\begin{equation}
\begin{aligned}\underset{\boldsymbol{\Sigma}\succ0}{\textrm{min}} & \textrm{Tr}\left[\boldsymbol{\Sigma}^{-1}\left(\mathbf{S}-\nu\mathbf{I}\right)\right]+\textrm{Tr}\left(\nu\boldsymbol{\Sigma}^{-1}\right)+\log\left|\boldsymbol{\Sigma}\right|\\
 & +\frac{\rho}{2}\left\Vert \boldsymbol{\Sigma}-P_{\mathcal{C}}\left(\boldsymbol{\Sigma}\right)\right\Vert _{F}^{2},
\end{aligned}
\end{equation}
where
\begin{equation}
\nu=\lambda_{\textrm{max}}\left(\mathbf{S}\right)\label{eq:49}
\end{equation}
Majorizing the distance penalty as in (\ref{eq:44-1}) and the concave
terms $\log\left|\boldsymbol{\Sigma}\right|$ and $\textrm{Tr}\left(\mathbf{\boldsymbol{\Sigma}}^{-1}\left(\mathbf{S}-\nu\mathbf{I}\right)\right)$
using their tangent hyperplanes, we get the following surrogate problem:
\begin{equation}
\begin{aligned}\underset{\boldsymbol{\Sigma}\succ0}{\textrm{min}} & -\textrm{Tr}\left[\left(\boldsymbol{\Sigma}^{t}\right)^{-1}\left(\mathbf{S}-\nu\mathbf{I}\right)\left(\boldsymbol{\Sigma}^{t}\right)^{-1}\boldsymbol{\Sigma}\right]+\textrm{Tr}\left(\left(\boldsymbol{\Sigma}^{t}\right)^{-1}\boldsymbol{\Sigma}\right)\\
 & +\textrm{Tr}\left(\nu\mathbf{\boldsymbol{\Sigma}}^{-1}\right)+\frac{\rho}{2}\left\Vert \boldsymbol{\Sigma}-P_{\mathcal{C}}\left(\boldsymbol{\Sigma}^{t}\right)\right\Vert _{F}^{2}
\end{aligned}
\end{equation}
 Expanding the distance penalty term and ignoring the constants yields
the following problem:
\begin{equation}
\begin{aligned}\underset{\boldsymbol{\Sigma}\succ0}{\textrm{min}} & \textrm{Tr}\left[\left(\left(\boldsymbol{\Sigma}^{t}\right)^{-1}-\left(\boldsymbol{\Sigma}^{t}\right)^{-1}\left(\mathbf{S}-\nu\mathbf{I}\right)\left(\boldsymbol{\Sigma}^{t}\right)^{-1}\right)\boldsymbol{\Sigma}\right]\\
 & +\textrm{Tr}\left(\nu\mathbf{\boldsymbol{\Sigma}}^{-1}\right)-\rho\textrm{Tr}\left[\boldsymbol{\Sigma}P_{\mathcal{C}}\left(\boldsymbol{\Sigma}^{t}\right)\right]+\frac{\rho}{2}\left\Vert \boldsymbol{\Sigma}\right\Vert _{F}^{2}
\end{aligned}
\end{equation}
which can be rewritten as:
\begin{equation}
\begin{aligned}\underset{\boldsymbol{\Sigma}\succ0}{\textrm{min}} & \textrm{Tr}\left\{ \left[\left(\boldsymbol{\Sigma}^{t}\right)^{-1}-\left(\boldsymbol{\Sigma}^{t}\right)^{-1}\left(\mathbf{S}-\nu\mathbf{I}\right)\left(\boldsymbol{\Sigma}^{t}\right)^{-1}\right.\right.\\
 & \left.\left.-\rho P_{\mathcal{C}}\left(\boldsymbol{\Sigma}^{t}\right)\right]\mathbf{\boldsymbol{\Sigma}}\right\} +\textrm{Tr}\left(\nu\boldsymbol{\Sigma}^{-1}\right)+\frac{\rho}{2}\left\Vert \boldsymbol{\Sigma}\right\Vert _{F}^{2}
\end{aligned}
\label{eq:5}
\end{equation}
Let
\begin{equation}
\mathbf{A}\triangleq\left(\boldsymbol{\Sigma}^{t}\right)^{-1}-\left(\boldsymbol{\Sigma}^{t}\right)^{-1}\left(\mathbf{S}-\nu\mathbf{I}\right)\left(\boldsymbol{\Sigma}^{t}\right)^{-1}-\rho P_{\mathcal{C}}\left(\boldsymbol{\Sigma}^{t}\right),\label{eq:A}
\end{equation}
then (\ref{eq:5}) can be compactly written as:
\begin{equation}
\begin{aligned}\underset{\boldsymbol{\Sigma}\succ0}{\textrm{min}}\; & \textrm{Tr}\left(\mathbf{A}\mathbf{\boldsymbol{\Sigma}}\right)+\textrm{Tr}\left(\nu\boldsymbol{\Sigma}^{-1}\right)+\frac{\rho}{2}\left\Vert \boldsymbol{\Sigma}\right\Vert _{F}^{2}\end{aligned}
\label{eq:46}
\end{equation}
The optimal solution of the above problem can be computed easily based
on the eigenvalue decomposition of the matrix $\mathbf{A}$. Let $\mathbf{V}$
and $\mathbf{E}\triangleq\textrm{diag}\left(\left[e_{1},e_{2},\ldots,e_{p}\right]\right)$
denote the matrices containing the eigenvectors and eigenvalues of
$\mathbf{A}$ such that $\mathbf{A}=\mathbf{V}\mathbf{E}\mathbf{V}^{T}$
and $e_{1}\geq e_{2}\geq\ldots\geq e_{p}$, and let $\mathbf{\mathbf{\Lambda}}\triangleq\textrm{diag}\left(\left[\lambda_{1},\lambda_{2},\ldots,\lambda_{p}\right]\right)$
denote the eigenvalues of $\mathbf{\boldsymbol{\Sigma}}$ such that
$\lambda_{1}\leq\lambda_{2}\leq\ldots\leq\lambda_{p}$. Then, by Ruhe's
trace inequality \cite{ruhe1970perturbation} (see also \cite{FAAN}):
\begin{equation}
\textrm{Tr}\left(\mathbf{A}\mathbf{\boldsymbol{\Sigma}}\right)\geq\stackrel[j=1]{p}{\sum}e_{j}\lambda_{j}\label{eq:52-1}
\end{equation}
It follows from (\ref{eq:46}) and (\ref{eq:52-1}) that the minimizer
of the problem (\ref{eq:46}) is $\mathbf{\boldsymbol{\Sigma}}=\mathbf{V}\mathbf{\mathbf{\Lambda}}\mathbf{V}^{T}$,
where $\left\{ \lambda_{i}\right\} $ are given by the solution to
the following constrained minimization problem:
\begin{equation}
\begin{aligned}\underset{\lambda_{i}>0}{\textrm{min}} & \stackrel[i=1]{p}{\sum}\left(\frac{\nu}{\lambda_{i}}+\lambda_{i}e_{i}+\frac{\rho}{2}\lambda_{i}^{2}\right)\\
\textrm{s.t. } & \lambda_{1}\leq\lambda_{2}\leq\ldots\leq\lambda_{p}
\end{aligned}
\label{6-1}
\end{equation}
To solve (\ref{6-1}), let us first consider the problem (\ref{6-1})
without constraints:
\begin{equation}
\underset{\lambda_{i}>0}{\textrm{min}}\stackrel[i=1]{p}{\sum}\left(\frac{\nu}{\lambda_{i}}+\lambda_{i}e_{i}+\frac{\rho}{2}\lambda_{i}^{2}\right),\label{eq:47}
\end{equation}
which is separable in the variables $\left\{ \lambda_{i}\right\} $.
Equating the derivative of (\ref{eq:47}) with respect to $\lambda_{i}$
to zero, we get:
\begin{equation}
\rho\lambda_{i}^{3}+e_{i}\lambda_{i}^{2}-\nu=0\label{eq:48}
\end{equation}
It follows from the Descartes rule of signs that (\ref{eq:48}) has
only one positive root. We next show that the positive roots $\left\{ \lambda_{i}\right\} $
of (\ref{eq:48}) (for $i=1,\ldots,p$) satisfy the constraint in
(\ref{6-1}) and therefore are the solutions of the problem (\ref{6-1}).

Let $\lambda_{1}$ and $\lambda_{2}$ be the two positive roots of
(\ref{eq:48}) corresponding to $i=1$ and $2$: 
\begin{align}
\rho\lambda_{1}+e_{1}-\frac{\nu}{\lambda_{1}^{2}}=0 & \;\textrm{and}\label{eq:51}\\
\rho\lambda_{2}+e_{2}-\frac{\nu}{\lambda_{2}^{2}}=0\label{eq:52}
\end{align}
Subtracting (\ref{eq:51}) from (\ref{eq:52}), we get
\begin{equation}
(\lambda_{1}-\lambda_{2})\left(\rho+\frac{\nu(\lambda_{1}+\lambda_{2})}{\lambda_{1}^{2}\lambda_{2}^{2}}\right)=e_{2}-e_{1}
\end{equation}
Since the term $\left(\rho+\frac{\nu(\lambda_{1}+\lambda_{2})}{\lambda_{1}^{2}\lambda_{2}^{2}}\right)$
is positive, the fact that $e_{2}\leq e_{1}$ implies $\lambda_{1}\geq\lambda_{2}$
and hence the constraint in (\ref{6-1}) is satisfied. 

Algorithm \ref{alg:Pseudo-code-of-Proximal} summarizes the pseudo-code
of the PD method. 

\begin{algorithm}[H]
{\footnotesize{}\caption{\label{alg:Pseudo-code-of-Proximal}The PD Algorithm}
}{\footnotesize\par}

\textbf{Input}: $\mathbf{S}$, $\epsilon$ , $\zeta$

\textbf{Initialize}: $t=0$, $\mathbf{\boldsymbol{\Sigma}}^{0}$, 

\textbf{Iterate: }
\begin{itemize}
\item log$\rho=$$(t+1)$log$\zeta$
\item Compute $P_{\mathcal{C}}\left(\boldsymbol{\Sigma}^{t}\right)$ as
the orthogonal projection of $\boldsymbol{\Sigma}^{t}$ on $\mathcal{C}$
and $\nu$ and $\mathbf{A}$ from (\ref{eq:49}) and (\ref{eq:A}).
\item Perform the eigenvalue decomposition $\mathbf{A=\mathbf{V}E\mathbf{V}}^{T}$.
\item Solve the cubic equation (\ref{eq:48}) to obtain $\left\{ \lambda_{i}\right\} $
as its positive solution.
\item $\boldsymbol{\Sigma}^{t+1}=\mathbf{V}\mathbf{\mathbf{\Lambda}}\mathbf{V}^{T}$
\item $t=t+1$
\end{itemize}
\textbf{Stop} and return $\boldsymbol{\Sigma}^{t+1}$ if $\left\Vert \boldsymbol{\Sigma}^{t+1}-\boldsymbol{\Sigma}^{t}\right\Vert _{F}\slash\left\Vert \boldsymbol{\Sigma}^{t}\right\Vert _{F}<\epsilon$

\textbf{Output}: $\hat{\boldsymbol{\Sigma}}=\boldsymbol{\Sigma}^{t+1}$
at convergence.
\end{algorithm}

Several remarks regarding the proposed algorithms follow:
\begin{itemize}
\item Both the proposed PD method and the method in \cite{xu2022proximal}
use the proximal distance algorithm principle. However, they are significantly
different from each other. The surrogate function proposed in \cite{xu2022proximal}
is constructed using local quadratic approximations and is not an
actual upper bound on the likelihood function; therefore, a monotonic
decrease of the objective is not ensured. Moreover, the minimizer
of the approximate surrogate function is also not guaranteed to be
positive definite. To deal with these issues, \cite{xu2022proximal}
needs backtracking at each iteration. In contrast to this, the surrogate
function used in the PD method tightly upperbounds the objective function,
and the proposed algorithm monotonically decrease the objective and
yields a positive definite estimate of the covariance matrix at each
iteration. 
\item The proposed PD method, with only a change in the definition of the
projection operator $P_{\mathcal{C}}\left(\cdot\right)$, can also
be used to solve the proximal distance problem considered in \cite{xu2022proximal},
where $\mathcal{C}$ is the set of symmetric matrices with at most
$2k$ non-zero off-diagonal elements. In fact, PD is capable of solving
the MLE problem with a wide-variety of constraints on the elements
of the covariance matrix. 
\item The main computational burden for BCD is the calculation of the inverse
of the matrix $\boldsymbol{\Sigma}_{\mathcal{BB}}^{t}$, and for PD
is the eigenvalue decomposition of the matrix $\mathbf{A}$, both
of which are of the order $\mathcal{O}\left(p^{3}\right)$. The memory
requirement for both the methods is $\mathcal{O}\left(p^{2}\right)$. 
\item The convergence of BCD can be proved using the results in Sections
4 and 5 of \cite{tseng2001convergence}, whereas the convergence of
PD can be established using the results in Section 4.4 of \cite{xu2022proximal}. 
\item We initialize the BCD algorithm using a random positive definite matrix
satisfying the sparsity constraint and the PD algorithm with a diagonal
matrix made from the sample variances. In addition, we initialize
the value of $\zeta$ in the PD as $1+\varepsilon$, where $\varepsilon$
is tuned for optimal convergence.
\end{itemize}

\section{\label{sec:Numerical-Simulations}Numerical Simulations}

In this section we illustrate the performance of the proposed methodology
using both synthetically generated data and real-world data. All simulations
are done using MATLAB on a personal computer with 2.8 GHz Intel(R)
Core(TM) i7-1165G7 CPU with $16$GB RAM. 

\subsection{Synthetic data}

We generate a true sparse covariance matrix $\mathbf{\boldsymbol{\Sigma}}_{\textrm{true}}$
with the desired level of sparsity (defined as the percentage of zero
elements) and condition number using the sprandsym command in MATLAB.
Then we generate the data samples \textbf{$\mathbf{y}_{i}$}, $i=1,...,n,$
as $\mathbf{\boldsymbol{\Sigma}}_{\textrm{true}}^{\frac{1}{2}}\eta$,
where the elements of $\eta$ are independently drawn from $\mathcal{N}(0,1)$. 

Using the so-generated data we will compare the proposed methods based
on coordinate descent and proximal distance. We will also compare
the proposed methods with two state-of-the-art methods in terms of
the normalized root mean square error (NRMSE) and the Matthews correlation
coefficient (MCC). 
\begin{figure}[h]
\begin{centering}
\begin{tabular}{c}
\includegraphics[scale=0.4]{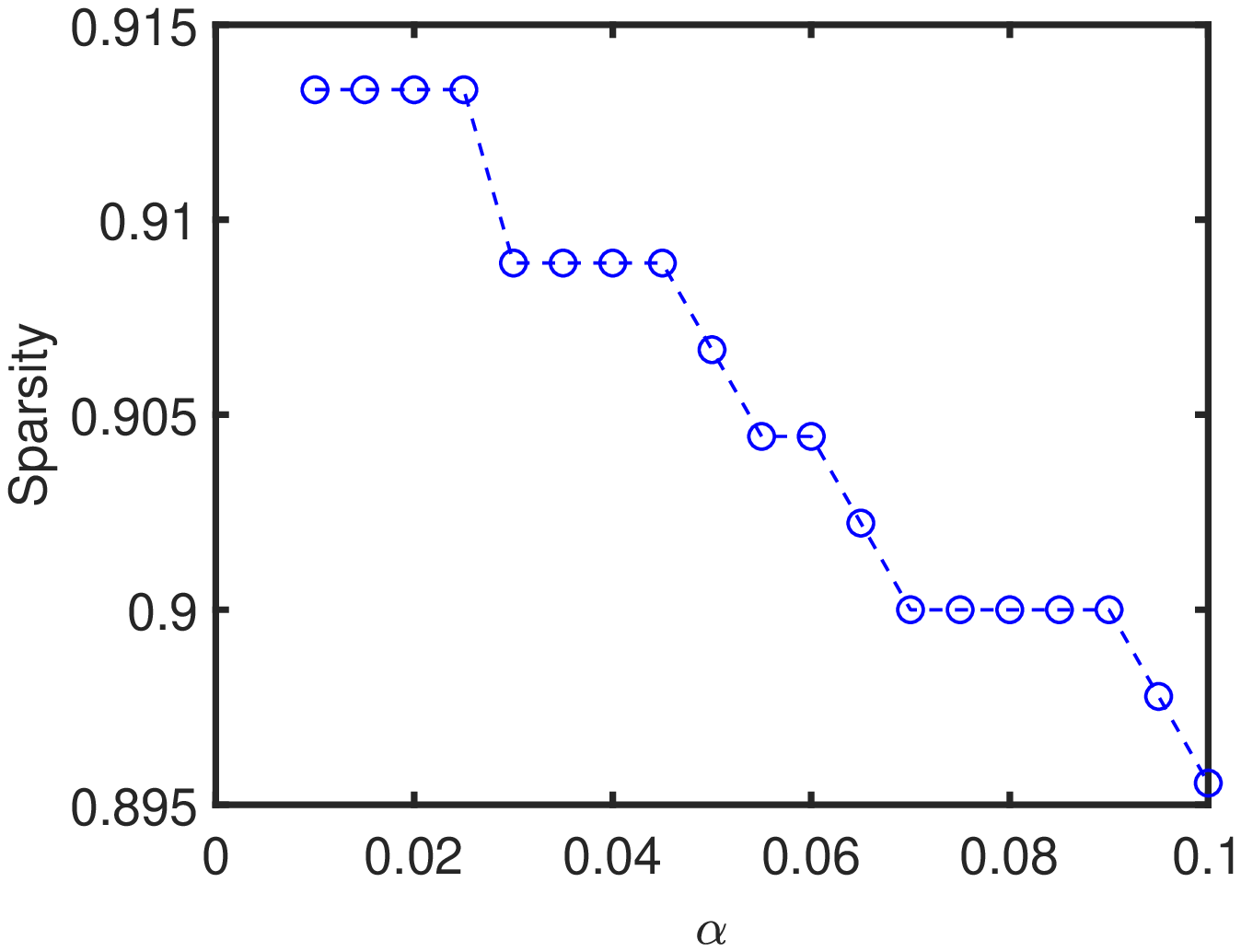}\tabularnewline
(a)\tabularnewline
\includegraphics[scale=0.4]{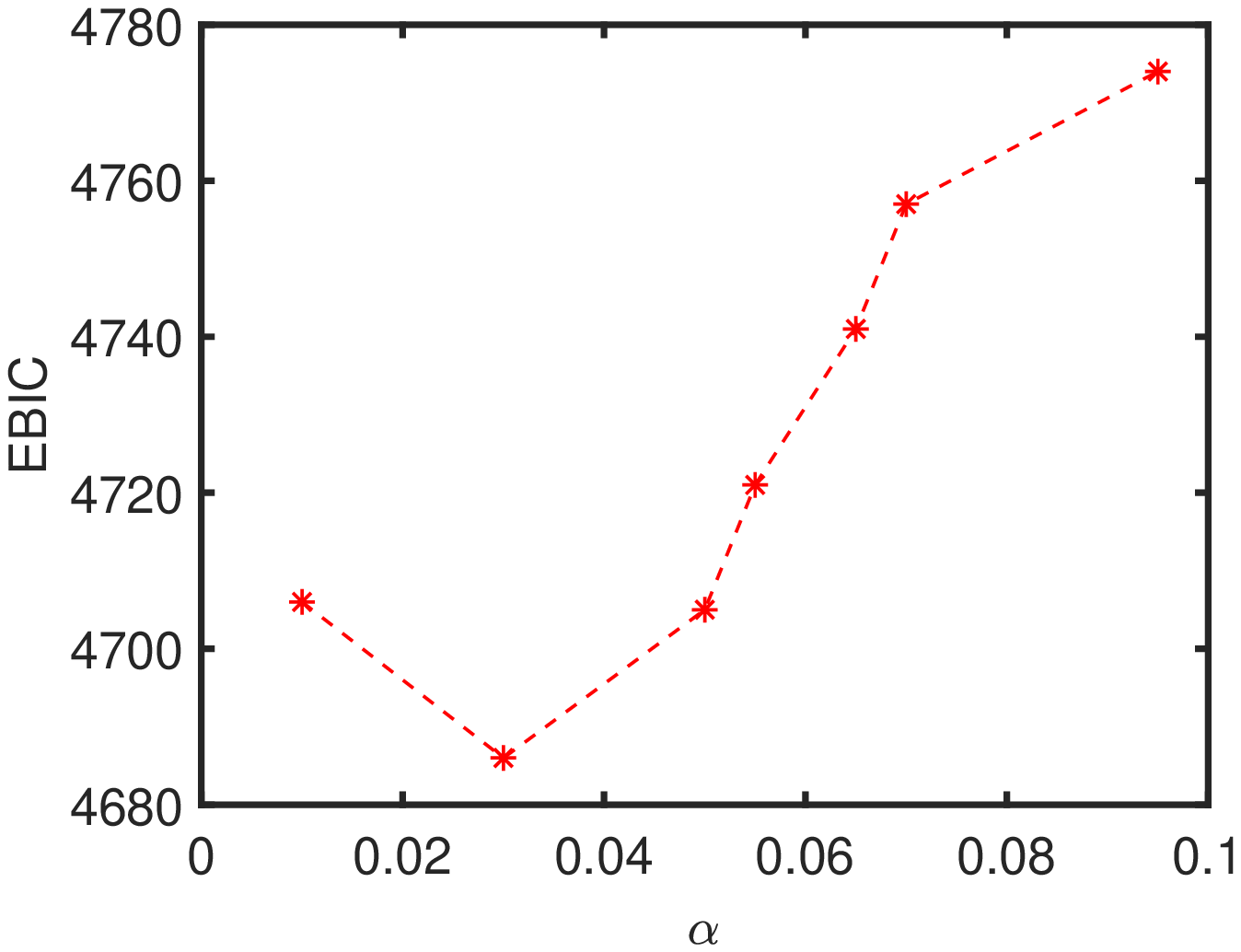}\tabularnewline
(b)\tabularnewline
\end{tabular}
\par\end{centering}
\caption{\label{fig:ebic}{\footnotesize{}(a) Variation of the sparsity obtained
with FDR for different values of $\alpha$ and (b) Variation of EBIC
for the selected values of $\alpha$ ($p=30$ , $n=40$ and $80\%$
sparsity).}}
\end{figure}
\begin{figure}[h]
\begin{centering}
\begin{tabular}{c}
\includegraphics[scale=0.4]{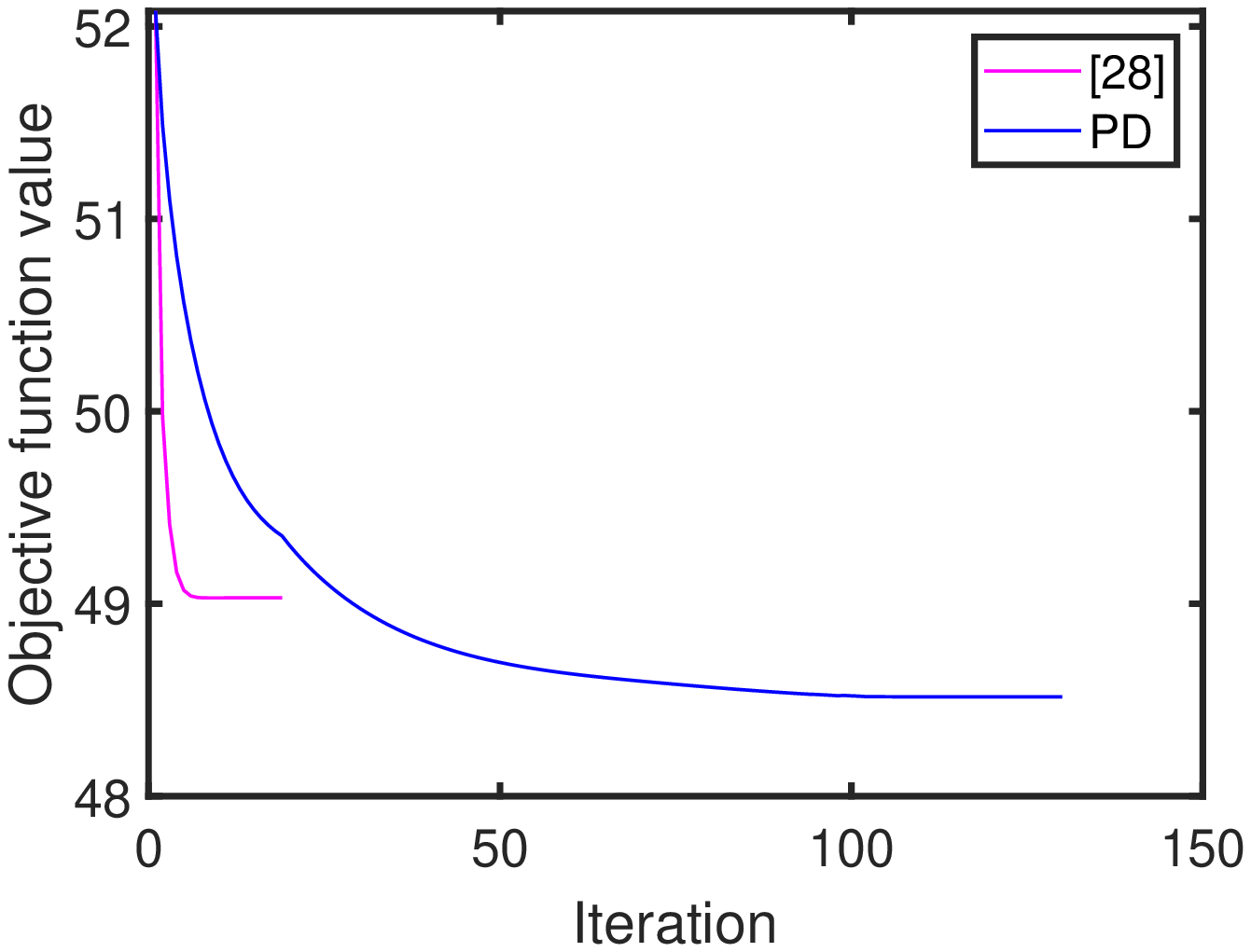}\tabularnewline
(a)\tabularnewline
\includegraphics[scale=0.4]{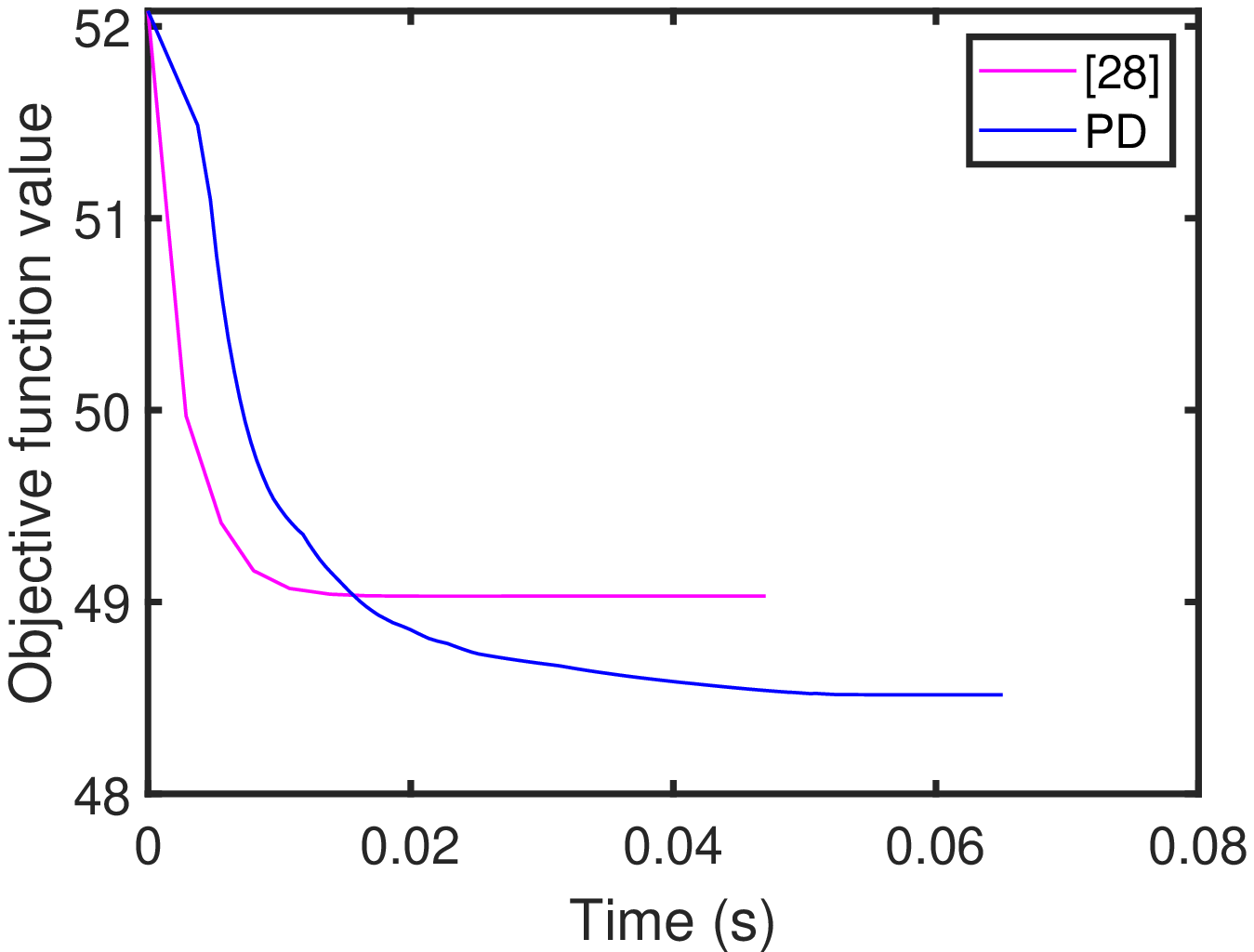}\tabularnewline
(b)\tabularnewline
\end{tabular}
\par\end{centering}
\caption{\label{fig:4}Variation of the objective function with (a) iteration
index and (b) Time (seconds) for $p=20$ and $n=40$ .}
\end{figure}

\subsubsection{Choosing $\alpha$}

To illustrate the choice of $\alpha$, we cover the interval $\left[0.01,0.1\right]$
using a fine grid with a step of $0.005$ and obtain the sparsity
corresponding to each grid point using the FDR. Fig. \ref{fig:ebic}(a)
shows the sparsity vs $\alpha$ plot for an example where $p=30$,
$n=40$ and the sparsity of the true covariance matrix is $80\%$.
From the figure it can be observed that the variation of sparsity
with $\alpha$ is step-wise. Also, the corresponding sparsity patterns
are hierarchical (see Section \ref{subsec:Choosing-the-optimal}).
In particular, this means that the same sparsity level implies the
same sparsity pattern, which allows us to evaluate EBIC only at a
small number of points in the interval, thereby reducing the computational
burden of choosing a value for $\alpha$. We pick one value of $\alpha$
for each sparsity level and evaluate EBIC at these points. Fig. \ref{fig:ebic}(b)
shows the variation of EBIC versus the selected values of $\alpha$
. The value of $\alpha$ that yields the lowest value of the criterion
is chosen. Note from Fig. \ref{fig:ebic}(b) that in the present case
we had to compute the MLE of $\boldsymbol{\Sigma}$ only for seven
values of $\alpha$ (i.e., seven sparsity patterns). 

\subsubsection{Comparison of PD with the algorithm proposed in \cite{xu2022proximal}}

Fig. \ref{fig:4}(a) shows the variation of the objective function
versus the iteration index for both methods. The two algorithms are
run on the same problem (\ref{eq:8}) for the choice of $k=40$. We
can see from the figure that the proposed PD method converges to a
lower value of the objective function than the method in \cite{xu2022proximal}
does. This difference can be attributed to the fact that the surrogate
in \cite{xu2022proximal} is not a real upper bound on the objective
function whereas the surrogate used in the PD tightly upperbounds
the objective function. The plot of the objective function versus
time is shown in Fig. \ref{fig:4}(b) for both methods. Note that
the time complexity of PD is comparable to that of the method in \cite{xu2022proximal}.
However the comparison in the latter figure does not include the time
needed for choosing the hyper-parameter for either method, which can
be potentially much larger for the method of \cite{xu2022proximal}
(as explained before). 

\subsubsection{BCD vs PD}

\begin{figure}[h]
\begin{centering}
\begin{tabular}{c}
\includegraphics[scale=0.4]{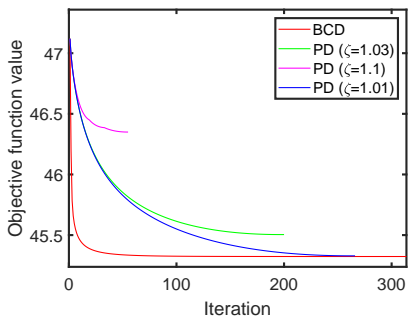}\tabularnewline
\end{tabular}
\par\end{centering}
\caption{\label{fig:3-1-2}{\footnotesize{}Objective function vs iteration
for $p=30$, $n=40$ and $\textrm{sparsity}=75\%$}}
\end{figure}

\begin{figure}[h]
\begin{centering}
\begin{tabular}{c}
\includegraphics[scale=0.4]{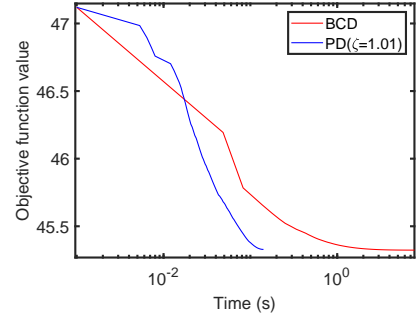}\tabularnewline
\end{tabular}
\par\end{centering}
\caption{\label{fig:3-1}{\footnotesize{}Objective function vs time (s) for
$p=30$, $n=40$ and $\textrm{sparsity}=75\%$}}
\end{figure}

Next we compare the performance of the two methods proposed in this
paper in terms of their computational complexity and convergence.
Fig. \ref{fig:3-1-2} shows the variation of the objective function
with the iteration index, whereas Fig. \ref{fig:3-1} shows the variation
versus time. From Fig. \ref{fig:3-1-2} we can see that PD converges
to the same solution as BCD if the penalty parameter is properly tuned.
However if $\zeta$ is not chosen correctly, the algorithm converges
to sub-optimal values. From Fig. \ref{fig:3-1} it can be observed
that PD with the a proper choice of $\zeta$ is not only capable of
converging to the same solution as BCD, but can be faster than BCD.
However, because the tuning of the penalty parameter $\zeta$ is not
simple, in the subsequent examples we will use only BCD.

\subsubsection{The effect of sparsity}

\begin{figure}[h]
\begin{centering}
\begin{tabular}{c}
\includegraphics[scale=0.4]{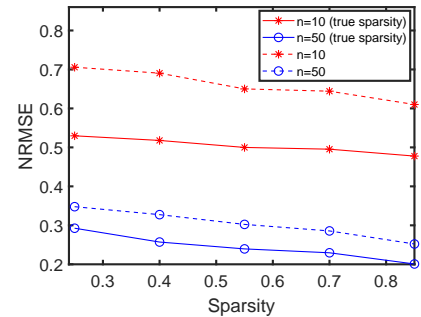}\tabularnewline
\end{tabular}
\par\end{centering}
\caption{\label{fig:3-1-1}{\footnotesize{}NRMSE vs sparsity for $p=30$. The
dashed curves show the NRMSE obtained with the sparsity pattern inferred
via FDR and the solid curves show the NRMSE obtained with the true
sparsity pattern.}}
\end{figure}

We study the effect of sparsity of the true covariance matrix on the
performance of BCD in terms of the normalized root mean square error
(NRMSE): 
\begin{equation}
\textrm{NRMSE}=\left\Vert \mathbf{\boldsymbol{\Sigma}}_{\textrm{true}}-\hat{\mathbf{\boldsymbol{\Sigma}}}\right\Vert _{F}\slash\left\Vert \mathbf{\boldsymbol{\Sigma}}_{\textrm{true}}\right\Vert _{F}.
\end{equation}
Fig. \ref{fig:3-1-1} shows the variation of NRMSE with the sparsity.
The dashed curves show the NRMSE obtained by solving the constrained
problem (\ref{eq:19-2}) with the matrix $\mathbf{Z}$ in the constraint
inferred via FDR (with $\alpha$ chosen via EBIC), whereas the solid
curves show the NRMSE obtained from (\ref{eq:19-2}) where $\mathbf{Z}$
corresponds to the sparsity pattern of $\mathbf{\boldsymbol{\Sigma}}_{\textrm{true}}$.
As expected the NRMSE is lower when $\mathbf{Z}$ encodes the true
sparsity pattern. However, the performance difference is not large
especially if $n>p$. Also, once again as expected, the NRMSE decreases
as sparsity of $\mathbf{\boldsymbol{\Sigma}}_{\textrm{true}}$ increases.

\subsubsection{Comparison with two state-of-the-art methods}

\begin{figure}[H]
\begin{centering}
\begin{tabular}{>{\centering}m{5cm}}
\multicolumn{1}{c}{\includegraphics[scale=0.4]{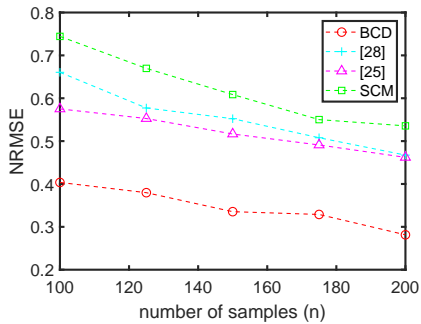}}\tabularnewline
\end{tabular}
\par\end{centering}
\caption{\label{fig:2}Variation of NRMSE with number of samples ($n)$ for
$p=150$ and sparsity $=66\%$.}
\end{figure}

Lastly, we compare BCD with SCM and the state-of-the-art methods of
\cite{bien2011sparse} and \cite{xu2022proximal}. Fig. \ref{fig:2}
shows the variation of NRMSE with the number of samples $n$ for $p=150$
and $n$ between $100$ and 2$00$. The true covariance matrix has
a sparsity of $66\%$. From the figure we can see that BCD yields
a lower value of NRMSE (with a difference of above $10\%$ or more)
for all values of $n$. 

\begin{figure}[h]
\begin{centering}
\includegraphics[scale=0.4]{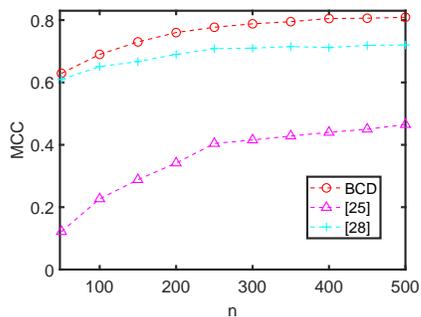}
\par\end{centering}
\caption{\label{fig:MCC-performance}MCC vs the number of samples ($n$) for
$p=30$, sparsity$=75\%$ and $n$ varying between $50$ and $500$.}
\end{figure}

We also use the Matthews correlation coefficient (MCC) to evaluate
the performance of the proposed algorithms w.r.t. recovering the true
covariance graph structure. MCC is defined as \cite{matthews1975comparison}:
\begin{equation}
\small\textrm{MCC}=\frac{TP.TN-FP.FN}{\sqrt{(TP+FP)(TP+FN)(TN+FP)(TN+FN)}},
\end{equation}
where $TP$, $TN$, $FP$, and $FN$ denote the number of true positives,
true negatives, false positives and false negatives, see the so-called
confusion matrix given in Table \ref{tab:Confusion-matrix}. A higher
value of MCC signifies fewer false alarms and misses, and therefore
a better performance. Fig. \ref{fig:MCC-performance} shows the variation
of MCC with the number of samples for $p=30$ and a true covariance
matrix with sparsity $75\%$. It can be seen that BCD has the highest
MCC scores. 

In both the experiments, the value of $\alpha$ for $\textrm{BCD}$
is estimated using EBIC, the regularization parameter $\lambda$ in
\cite{bien2011sparse} is chosen by a rather time consuming cross-validation
operation, whereas $k$ in \cite{xu2022proximal} is chosen equal
to the number of nonzero entries in the upper triangle of the true
covariance matrix (a choice that of course would not be feasible in
applications). 
\begin{table}[h]
\caption{\label{tab:Confusion-matrix}Confusion matrix}

\centering{}\resizebox{6cm}{!}{%
\begin{tabular}{|c|>{\centering}p{1.75cm}|>{\centering}p{1.5cm}|}
\hline 
\multirow{2}{*}{{\footnotesize{}\diagbox{Ground Truth}{Decision}}} & {\footnotesize{}$H_{ij}$ = accepted} & {\footnotesize{}$H_{ij}$ = rejected}\tabularnewline
 & {\footnotesize{}($\hat{\Sigma}_{ij}=0$)} & {\footnotesize{}($\hat{\Sigma}_{ij}\neq0$)}\tabularnewline
\hline 
\hline 
{\footnotesize{}$H_{ij}$ = true} & {\footnotesize{}$TN$} & {\footnotesize{}$FP$}\tabularnewline
{\footnotesize{}($\Sigma_{ij}=0$)} & {\footnotesize{}(detection)} & {\footnotesize{}(false alarm)}\tabularnewline
\hline 
{\footnotesize{}$H_{ij}$ = false} & {\footnotesize{}$FN$} & {\footnotesize{}$TP$}\tabularnewline
{\footnotesize{}($\Sigma_{ij}\neq0$)} & {\footnotesize{}(miss)} & {\footnotesize{}(detection)}\tabularnewline
\hline 
\end{tabular}}
\end{table}

\subsection{Real data}

To test the performance of BCD on real data we consider two datasets:
international migration forecast data that consist of the net migration
estimates in different countries and cell signalling data that consist
of the flow cytometry measurements of proteins in the cells. 
\begin{figure}[h]
\begin{centering}
\begin{tabular}{>{\centering}p{4.1cm}||>{\centering}p{4.1cm}>{\centering}p{4.1cm}||>{\centering}p{4cm}}
\multicolumn{4}{c}{\includegraphics[scale=0.3]{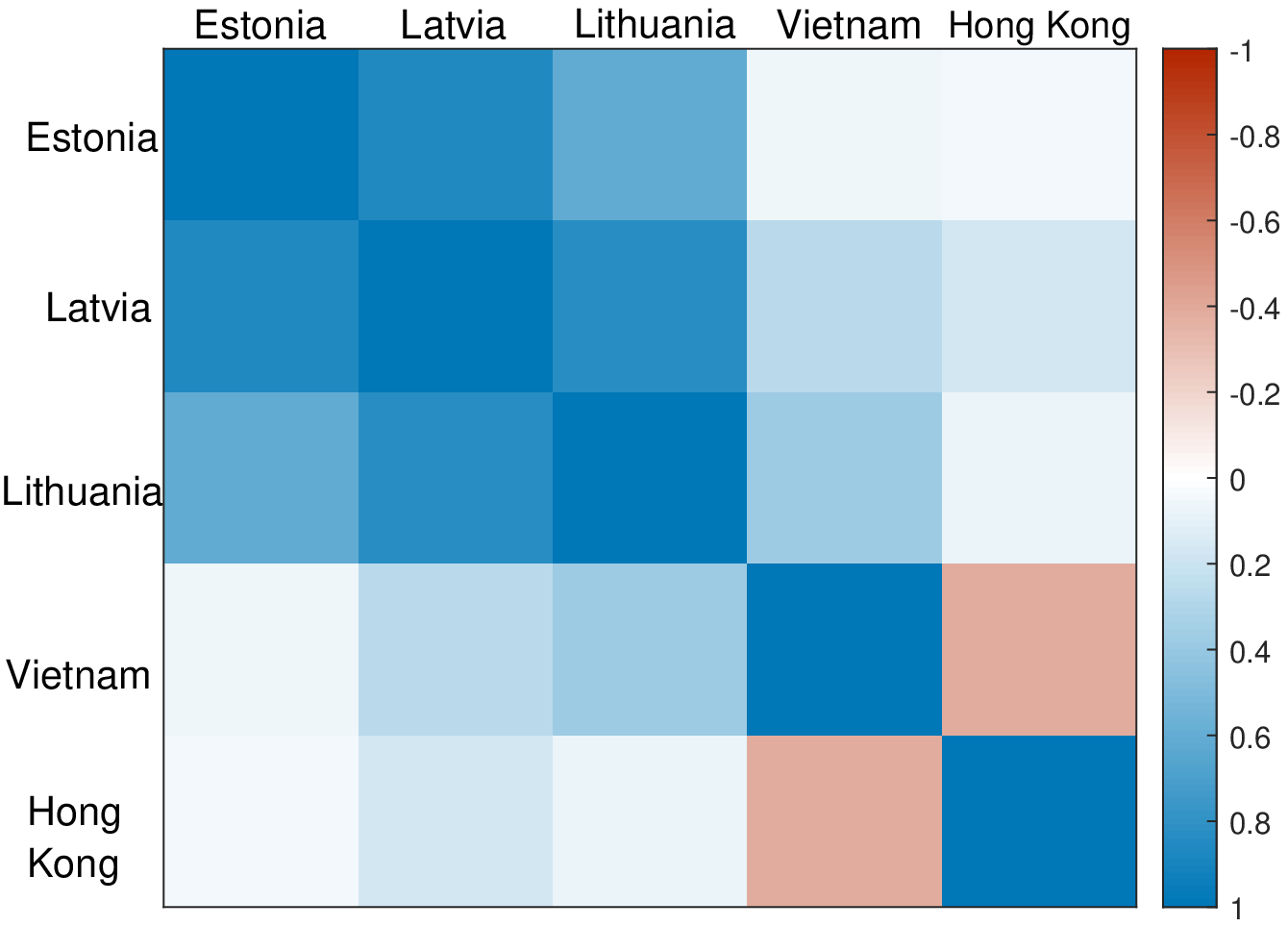}\includegraphics[scale=0.3]{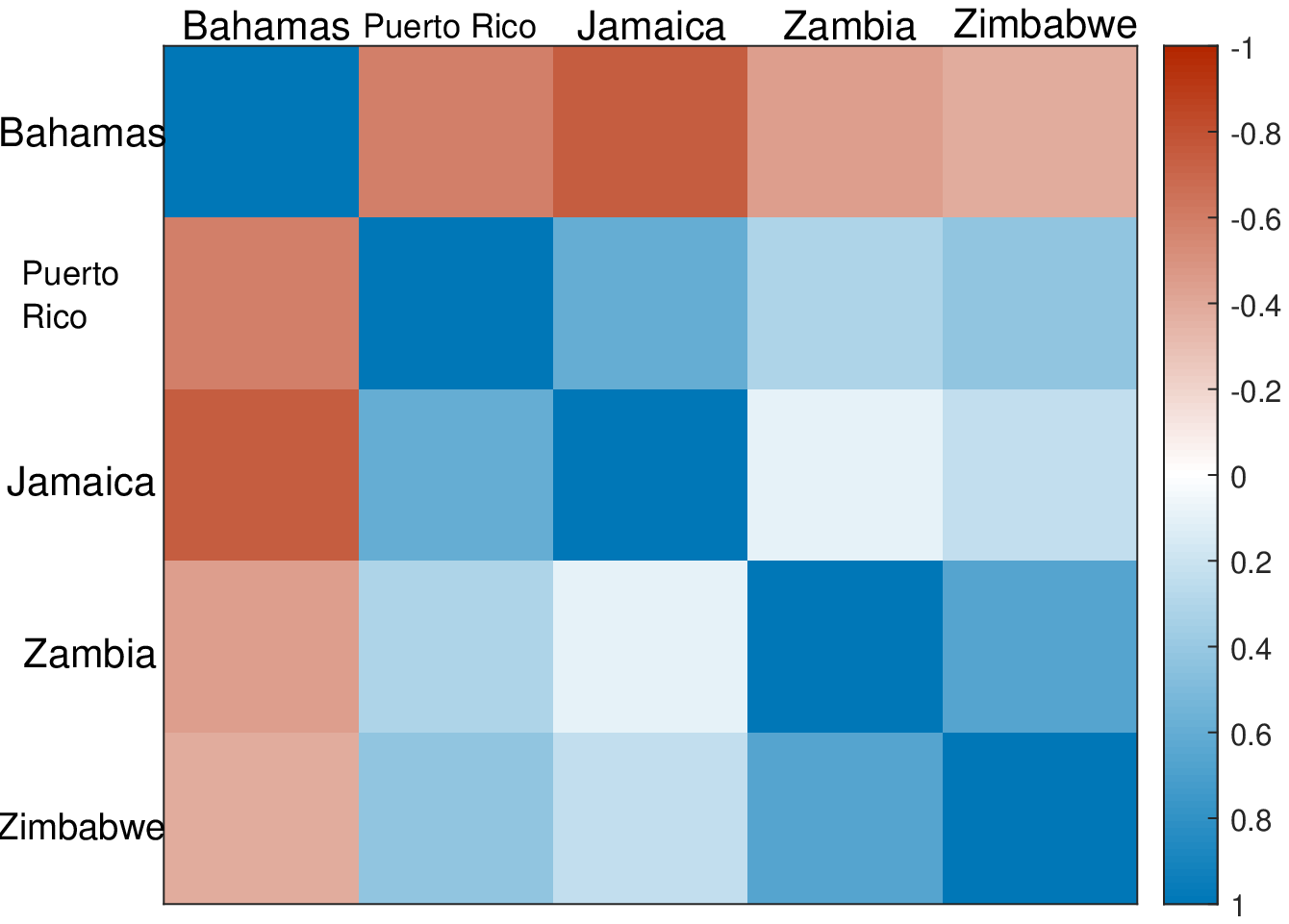}}\tabularnewline
\multicolumn{2}{>{\centering}p{4cm}}{(a)} & \multicolumn{2}{c}{(d)}\tabularnewline
\multicolumn{4}{c}{\includegraphics[scale=0.3]{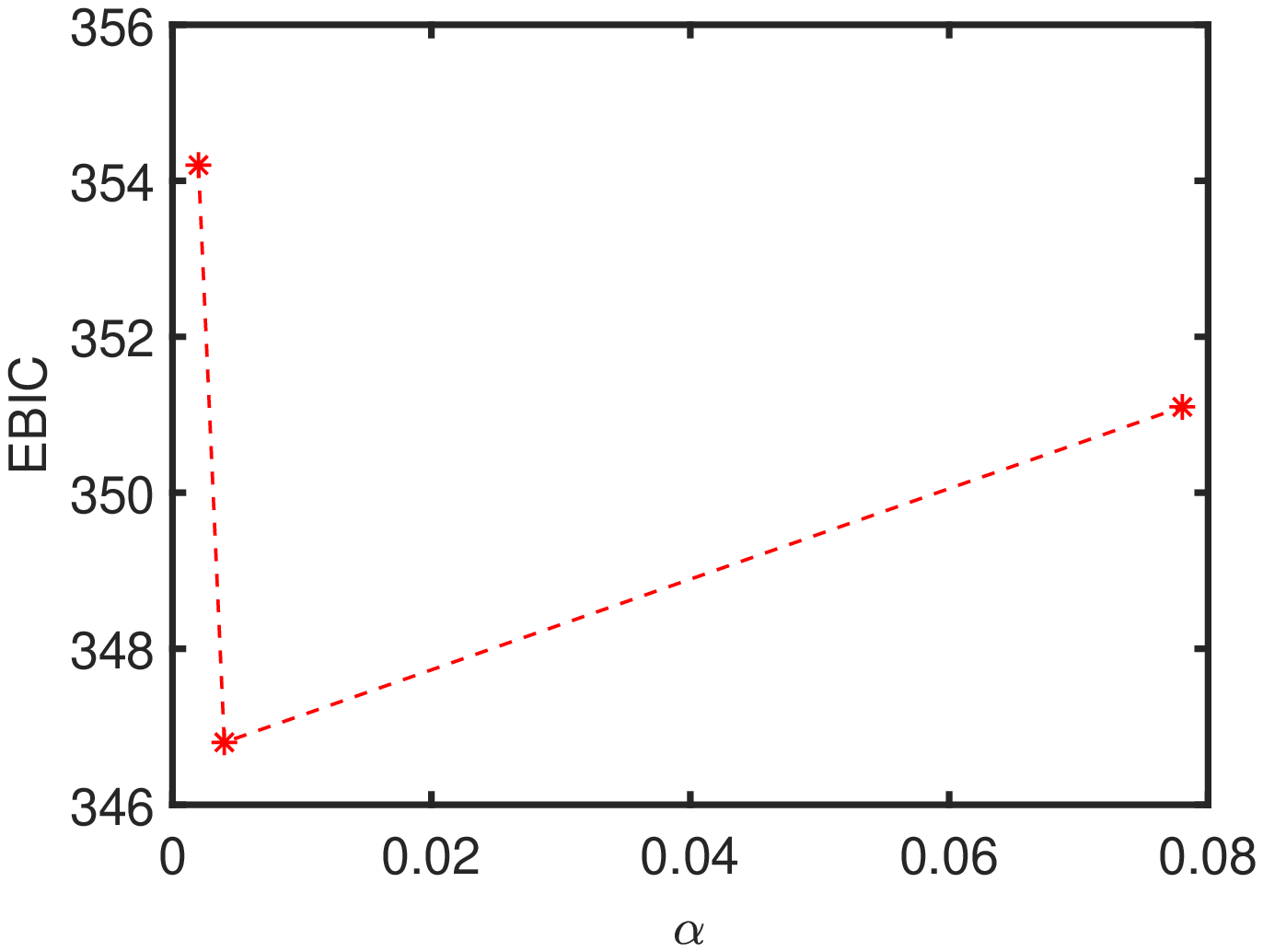}\includegraphics[scale=0.3]{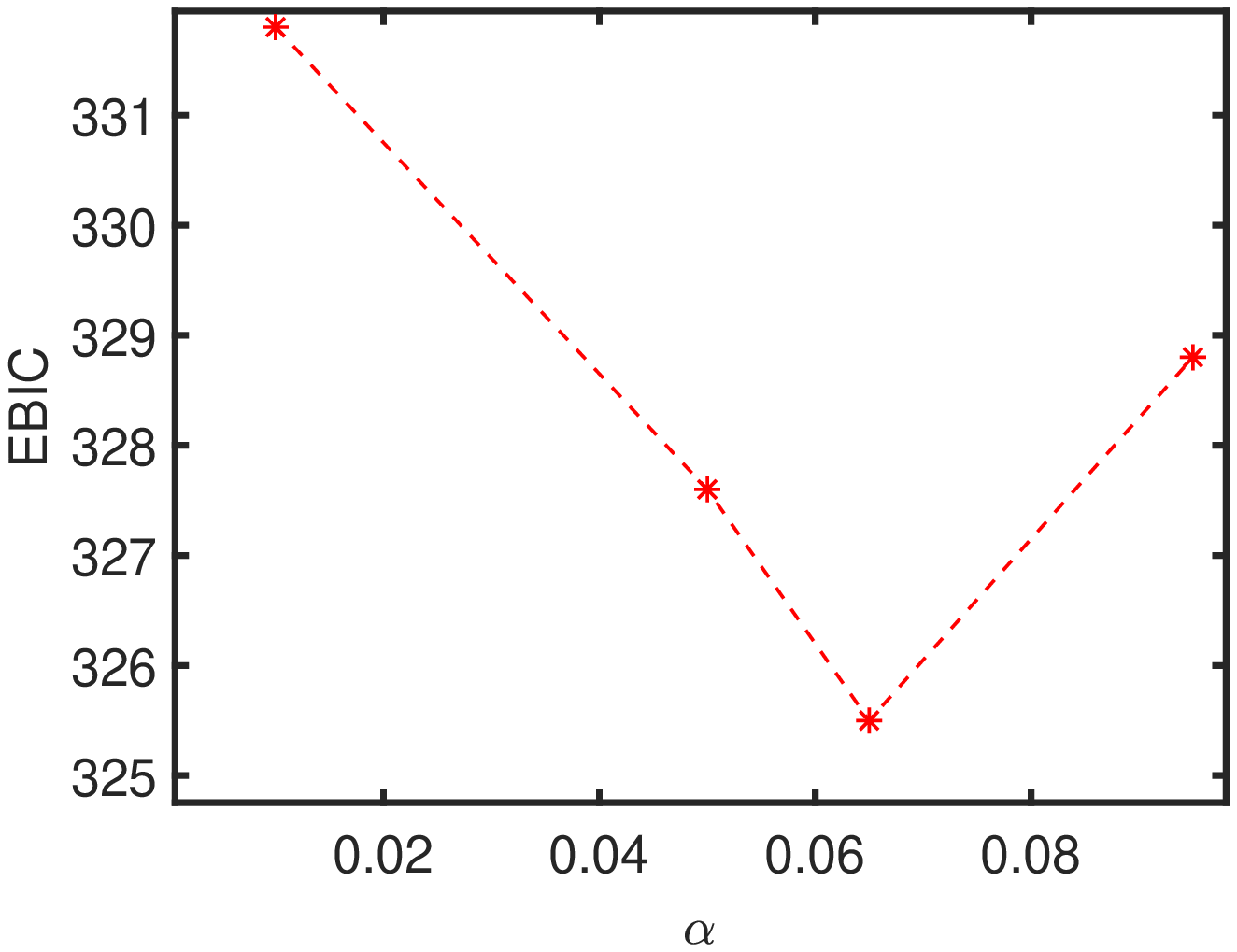}}\tabularnewline
\multicolumn{2}{>{\centering}p{4cm}}{(b)} & \multicolumn{2}{c}{(e)}\tabularnewline
\multicolumn{4}{c}{\includegraphics[scale=0.3]{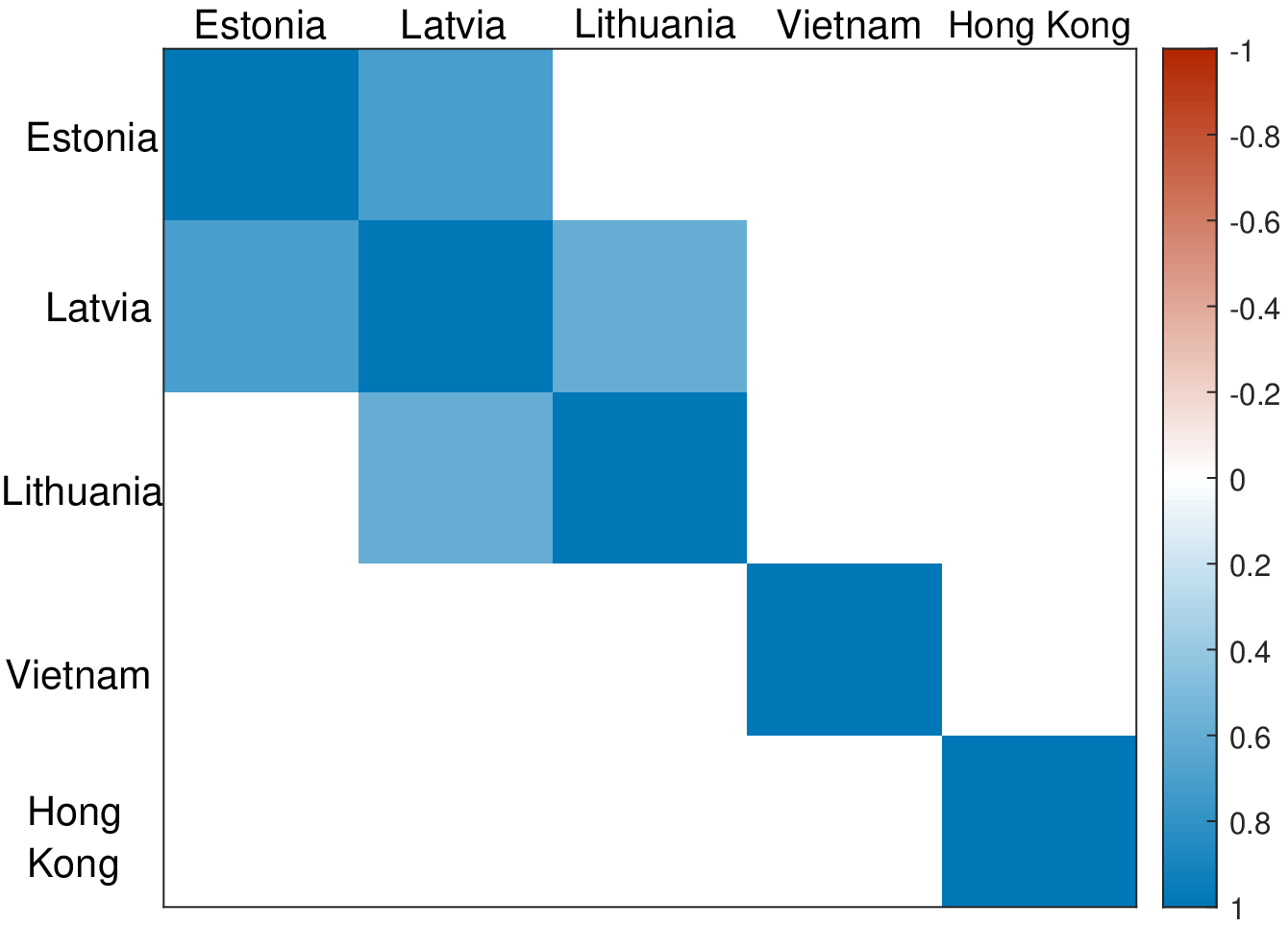}\includegraphics[scale=0.3]{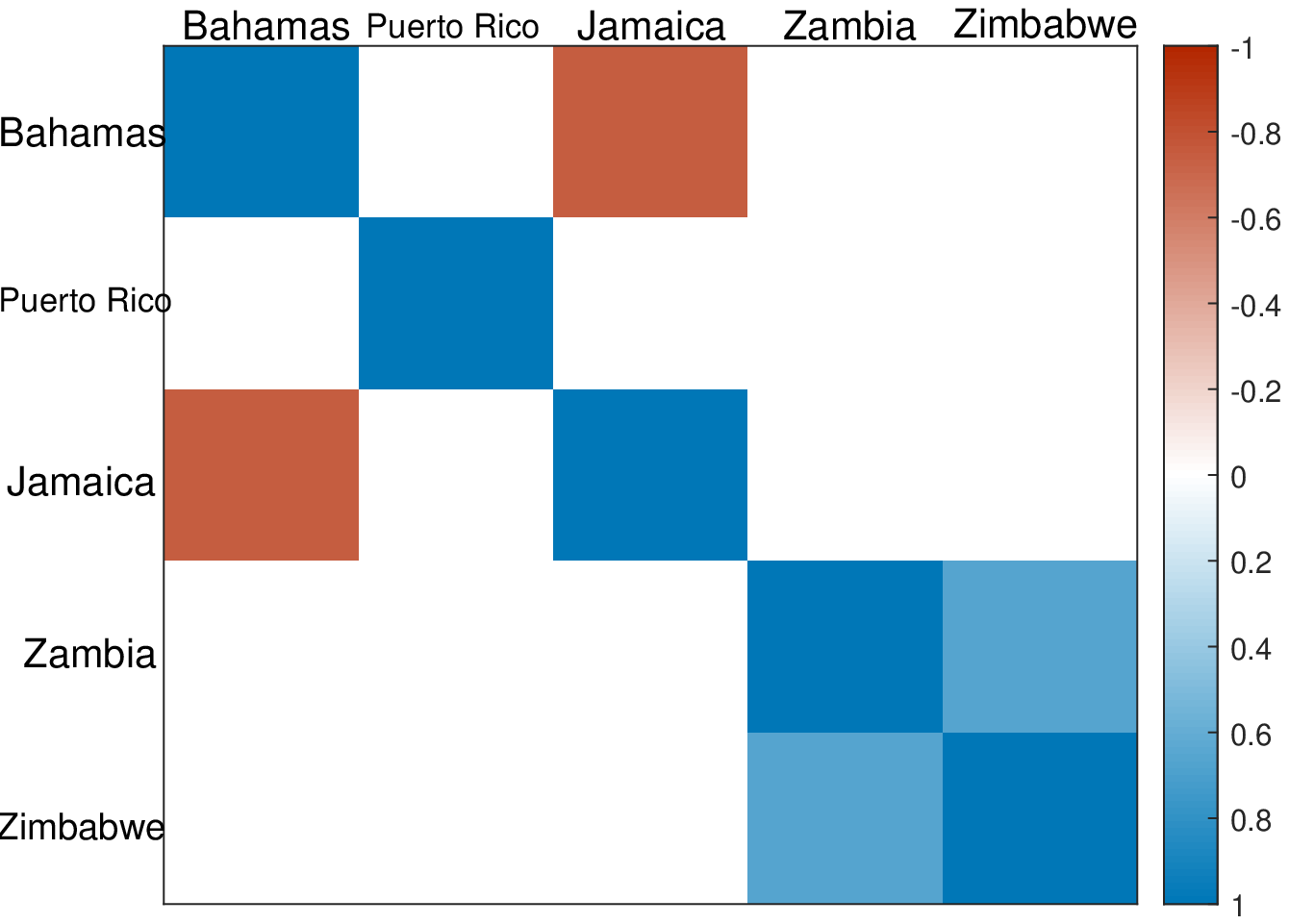}}\tabularnewline
\multicolumn{2}{>{\centering}p{4cm}}{(c)} & \multicolumn{2}{c}{(f)}\tabularnewline
\end{tabular}
\par\end{centering}
\caption{\label{fig:7}Two examples of estimated correlations of five countries.
Example 1 (left-hand side): (a) Pearson correlation estimates; (b)
EBIC vs $\alpha$; (c) Estimated correlations using BCD ($\alpha=0.002$).
Example 2 (right-hand side): (a) Pearson correlation estimates; (b)
EBIC vs $\alpha$; (c) Estimated correlations using BCD ($\alpha=0.065$)}
\end{figure}
\begin{figure*}[tbh]
\begin{centering}
\begin{tabular}{>{\centering}p{4.1cm}>{\centering}p{4.1cm}>{\centering}p{4.1cm}c||c}
\multicolumn{5}{c}{\includegraphics[scale=0.31]{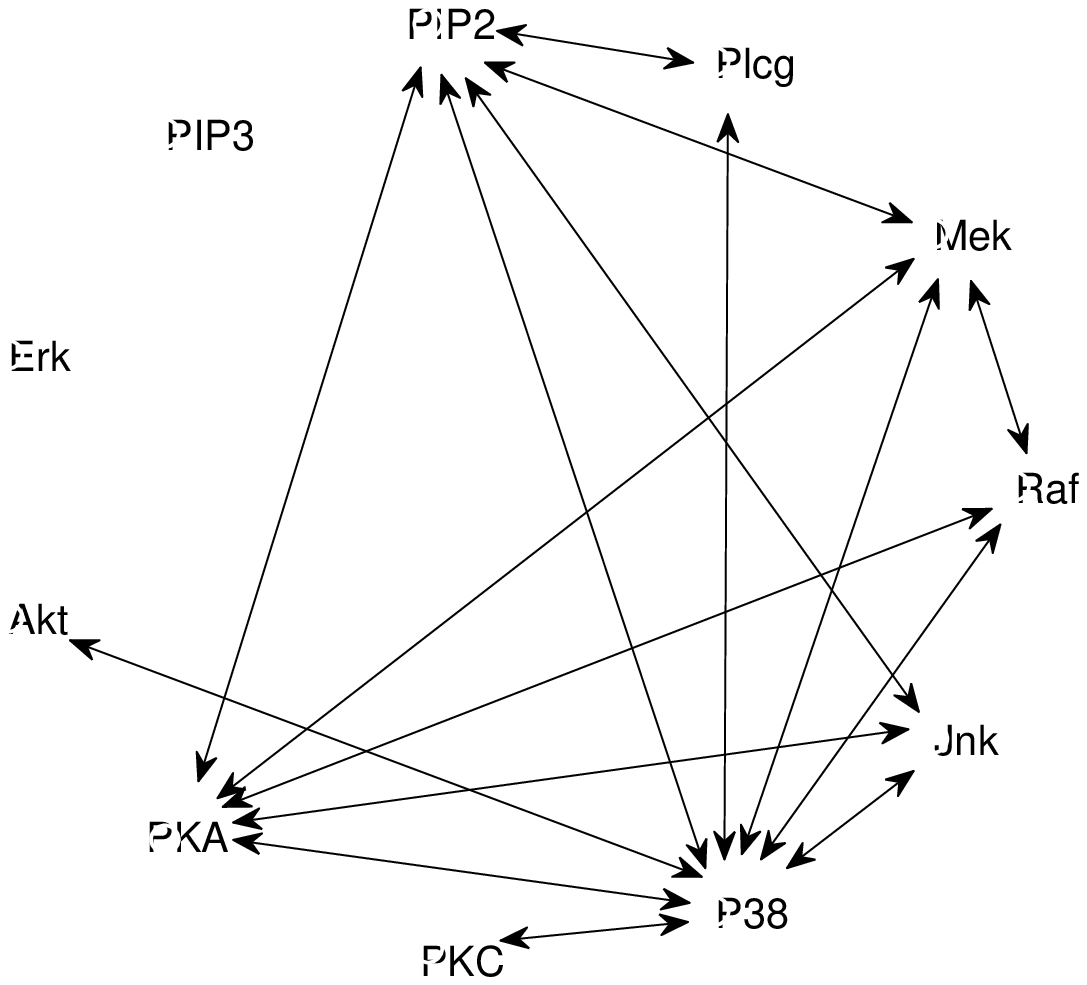}\includegraphics[scale=0.31]{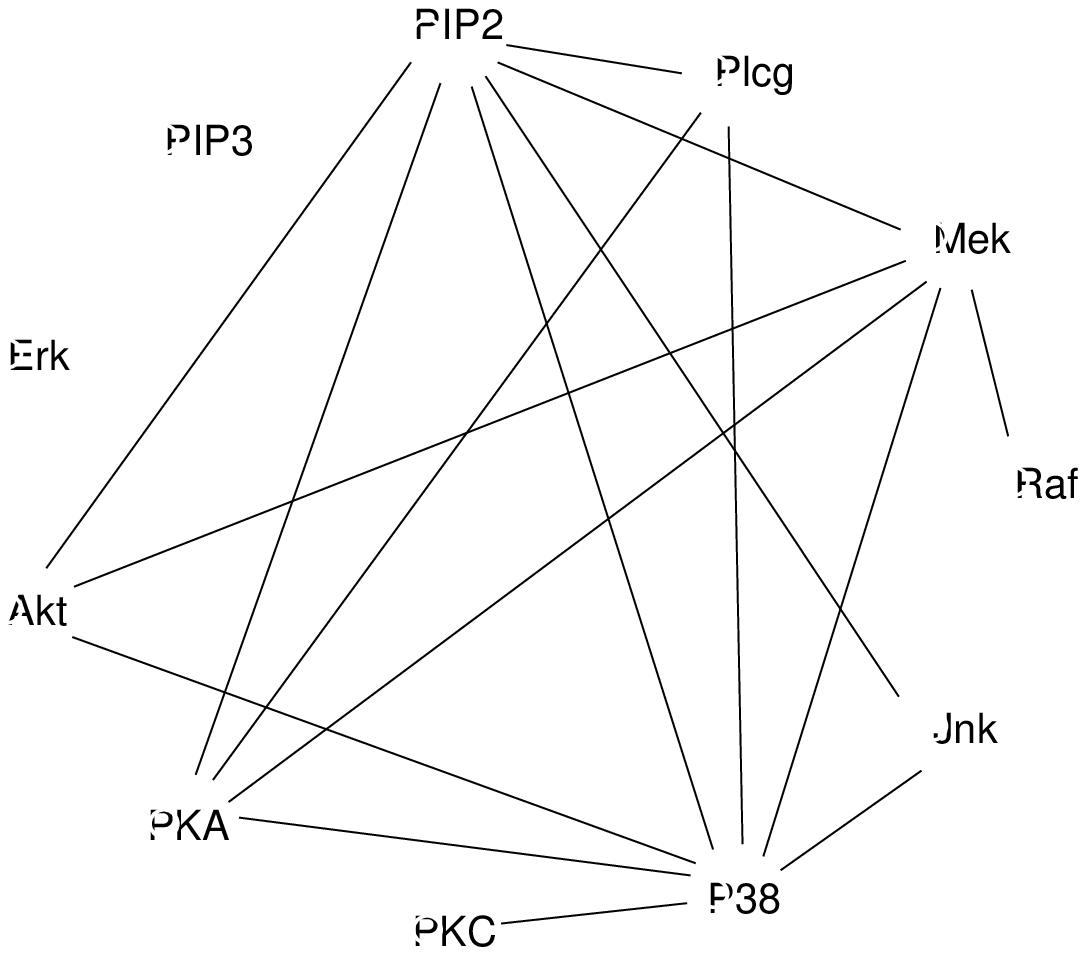}\includegraphics[scale=0.31]{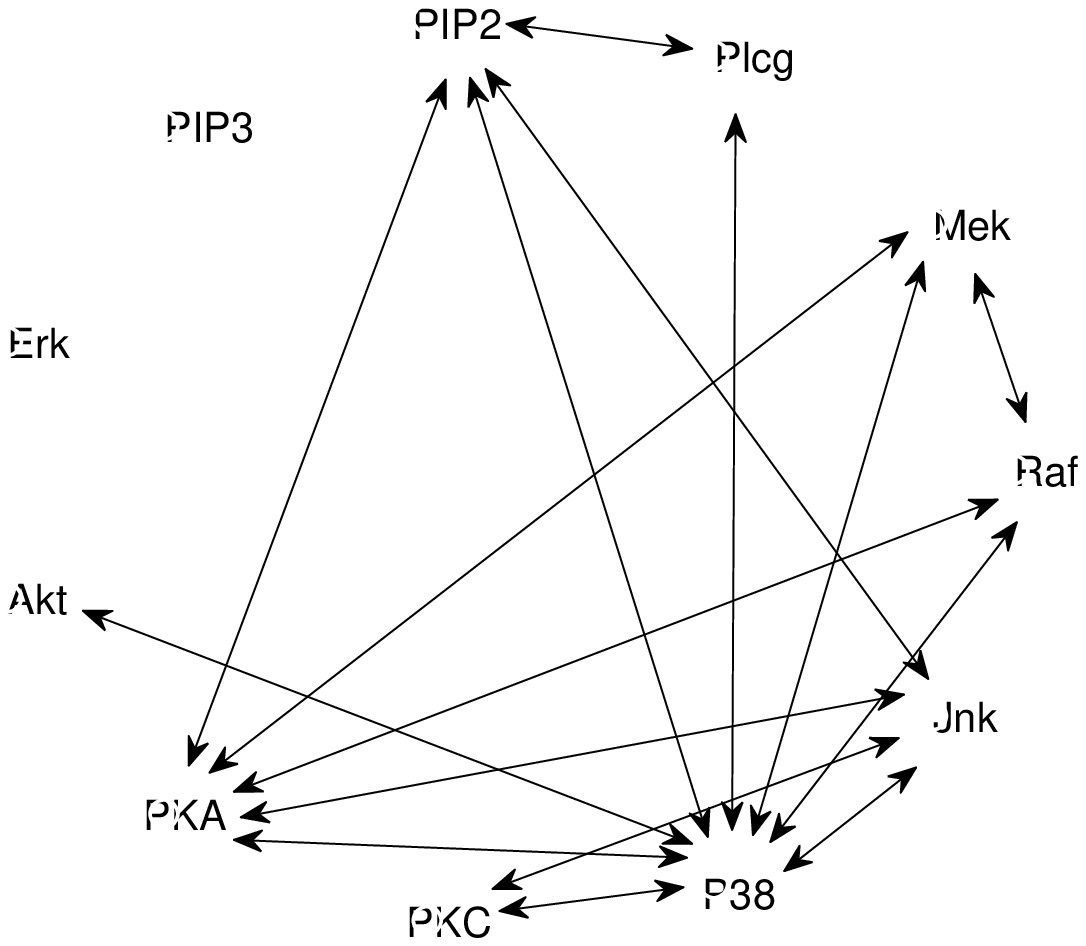}\includegraphics[scale=0.31]{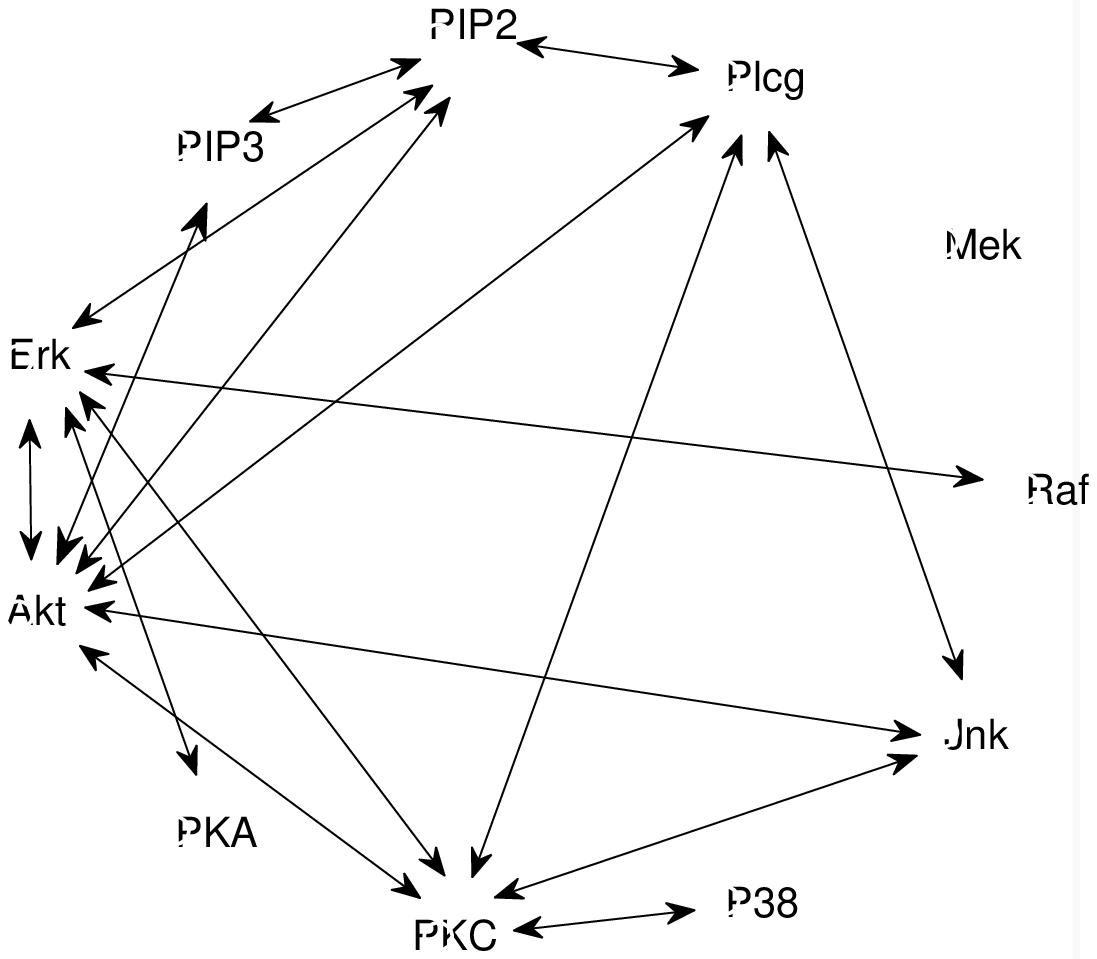}}\tabularnewline
\multicolumn{5}{c}{\includegraphics[scale=0.31]{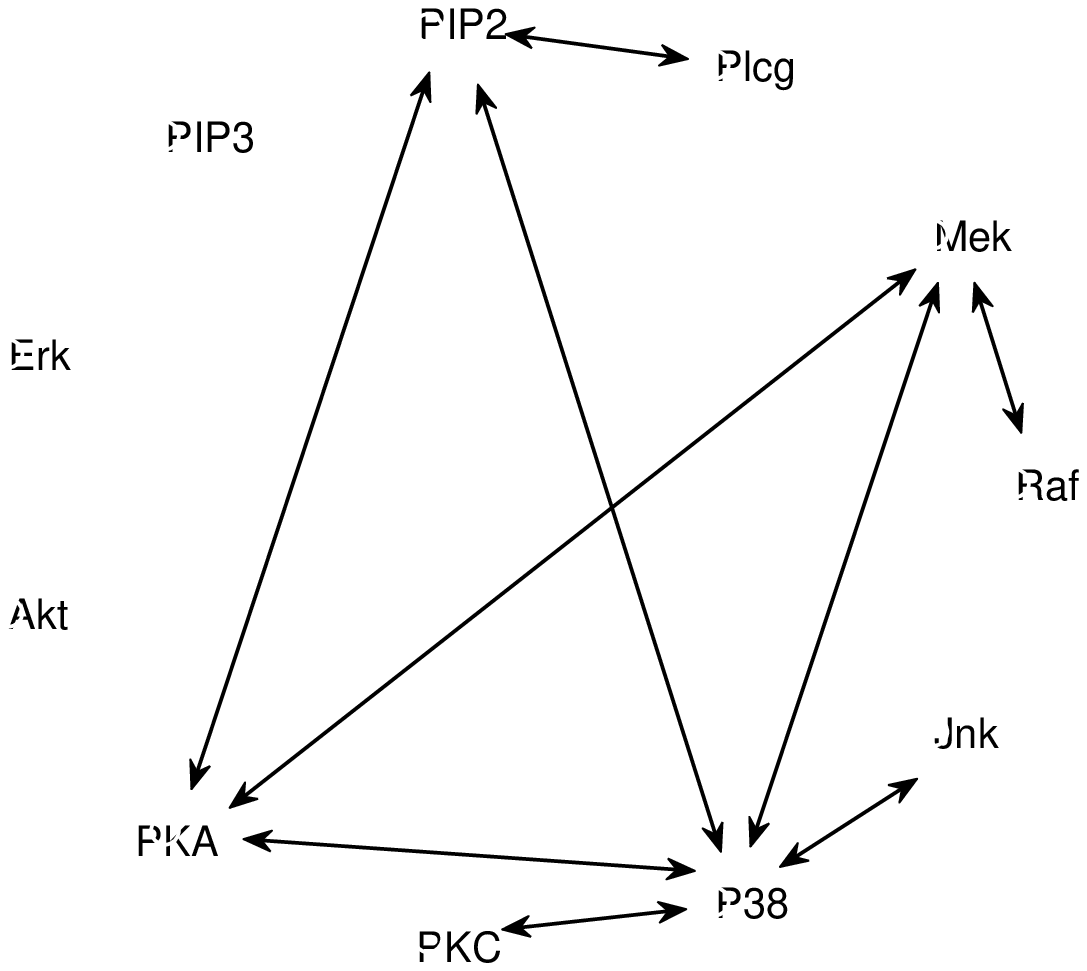}\includegraphics[scale=0.31]{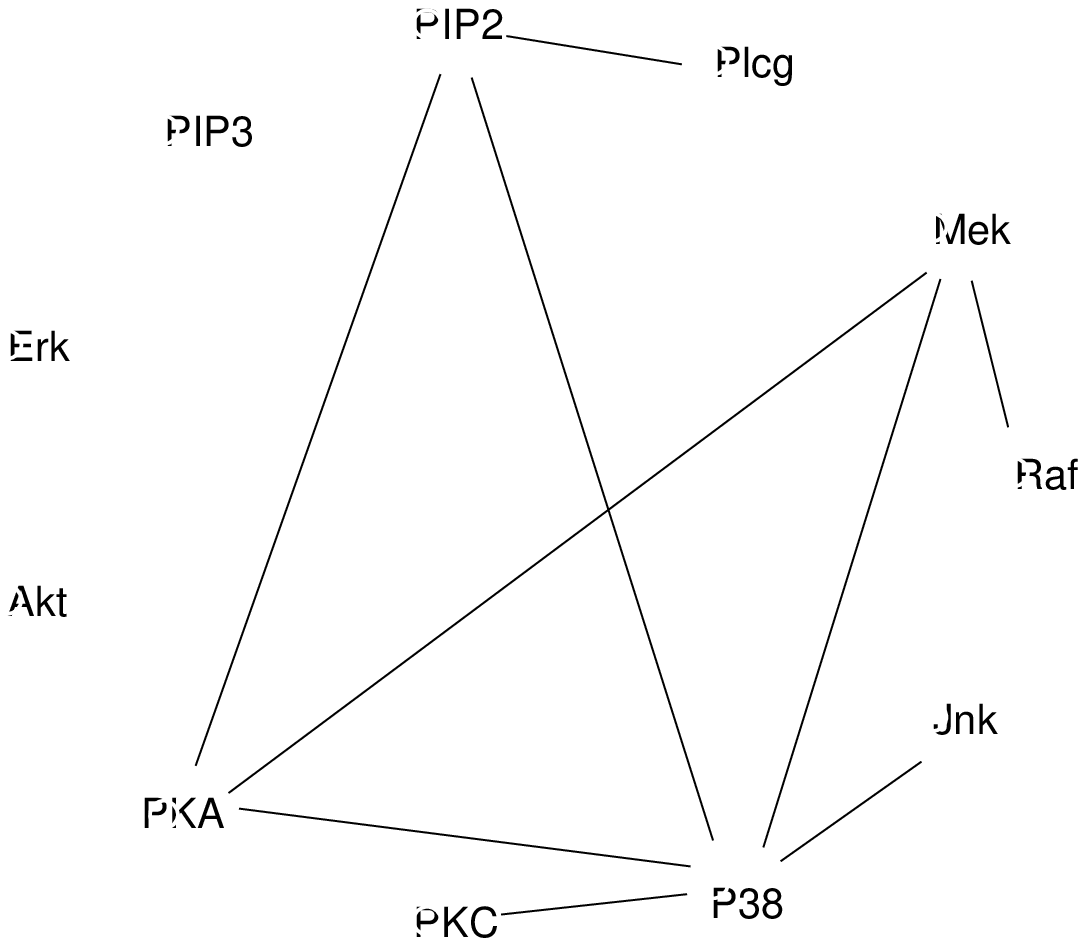}\includegraphics[scale=0.31]{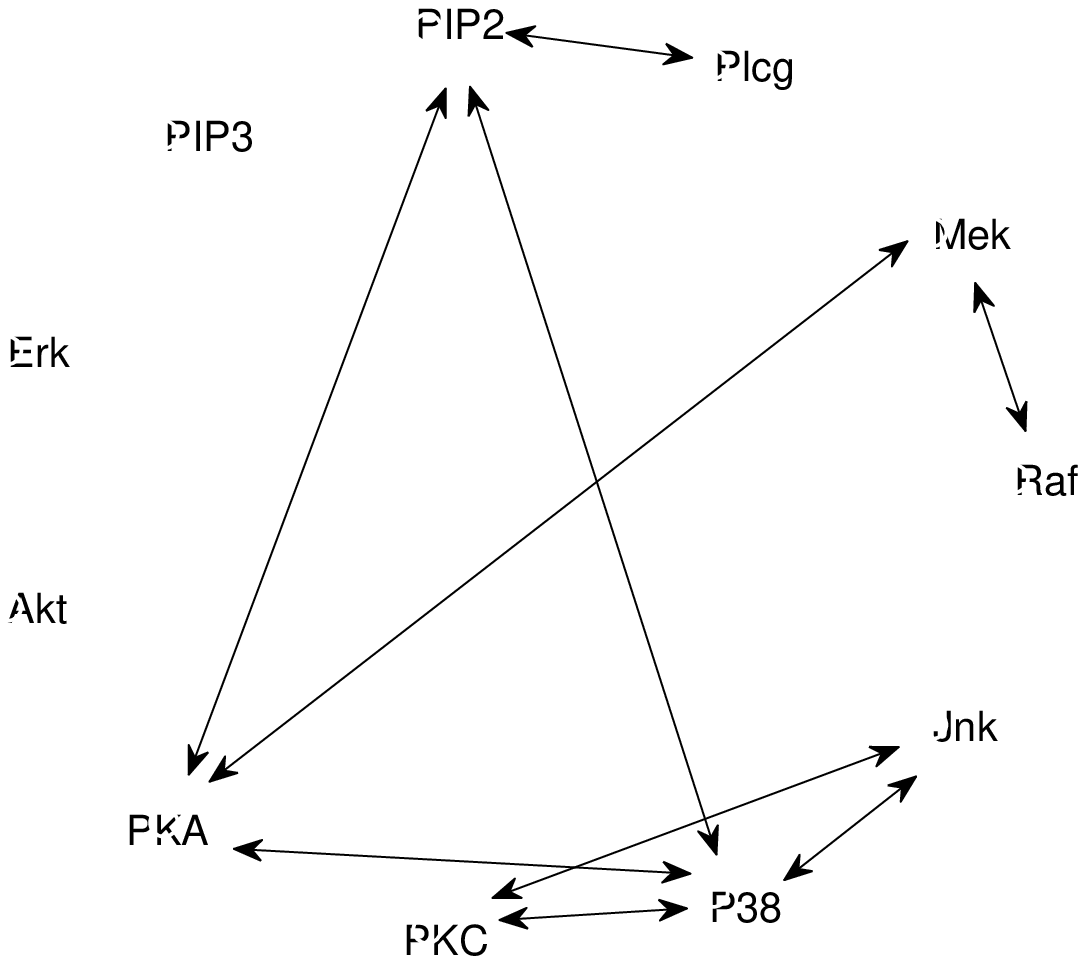}\includegraphics[scale=0.31]{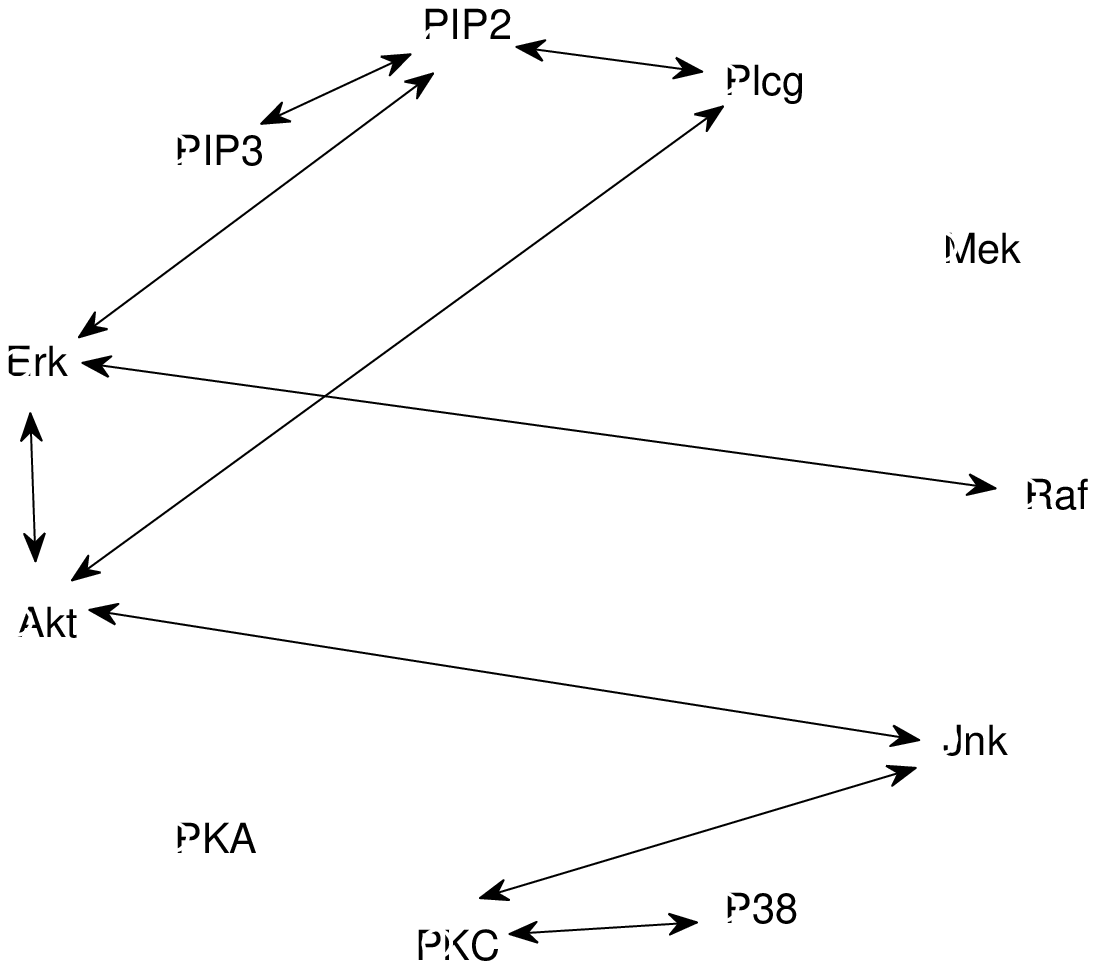}}\tabularnewline
\multicolumn{5}{c}{\includegraphics[scale=0.31]{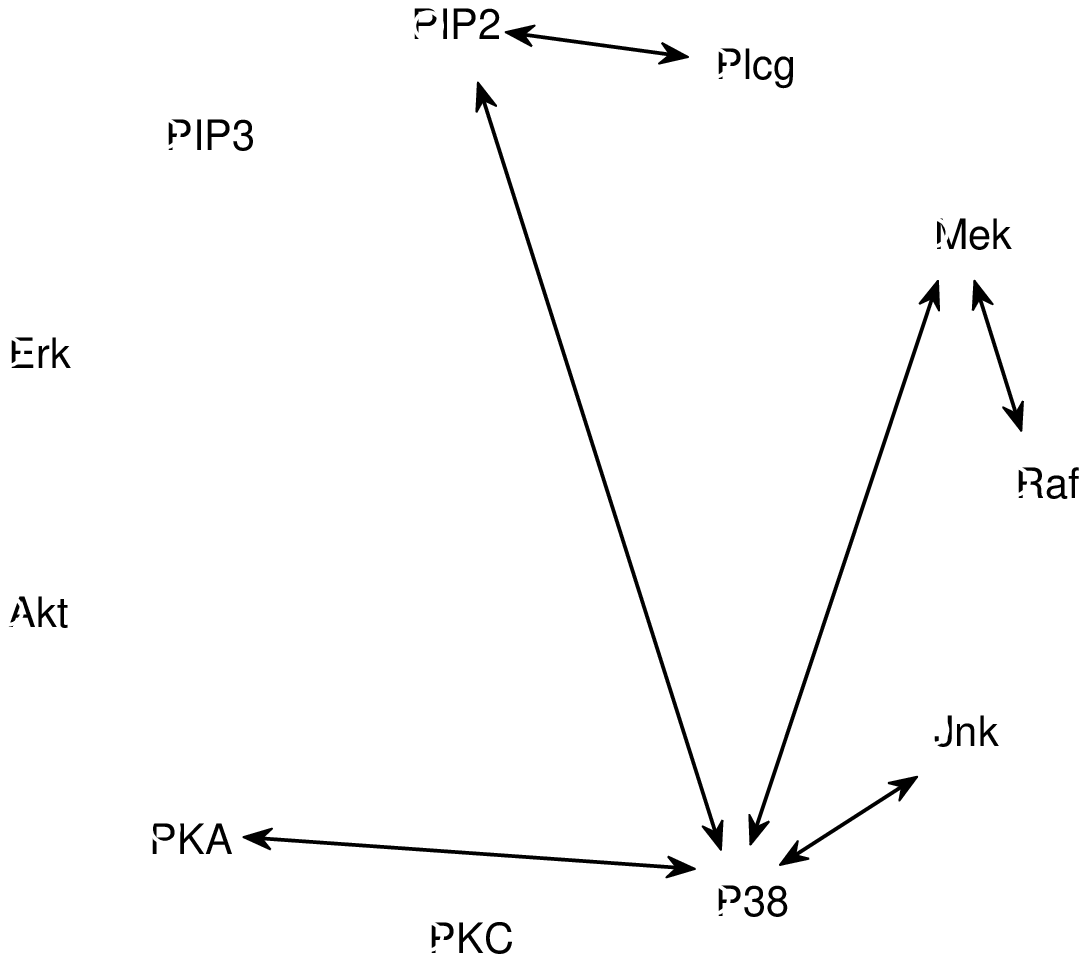}\includegraphics[scale=0.31]{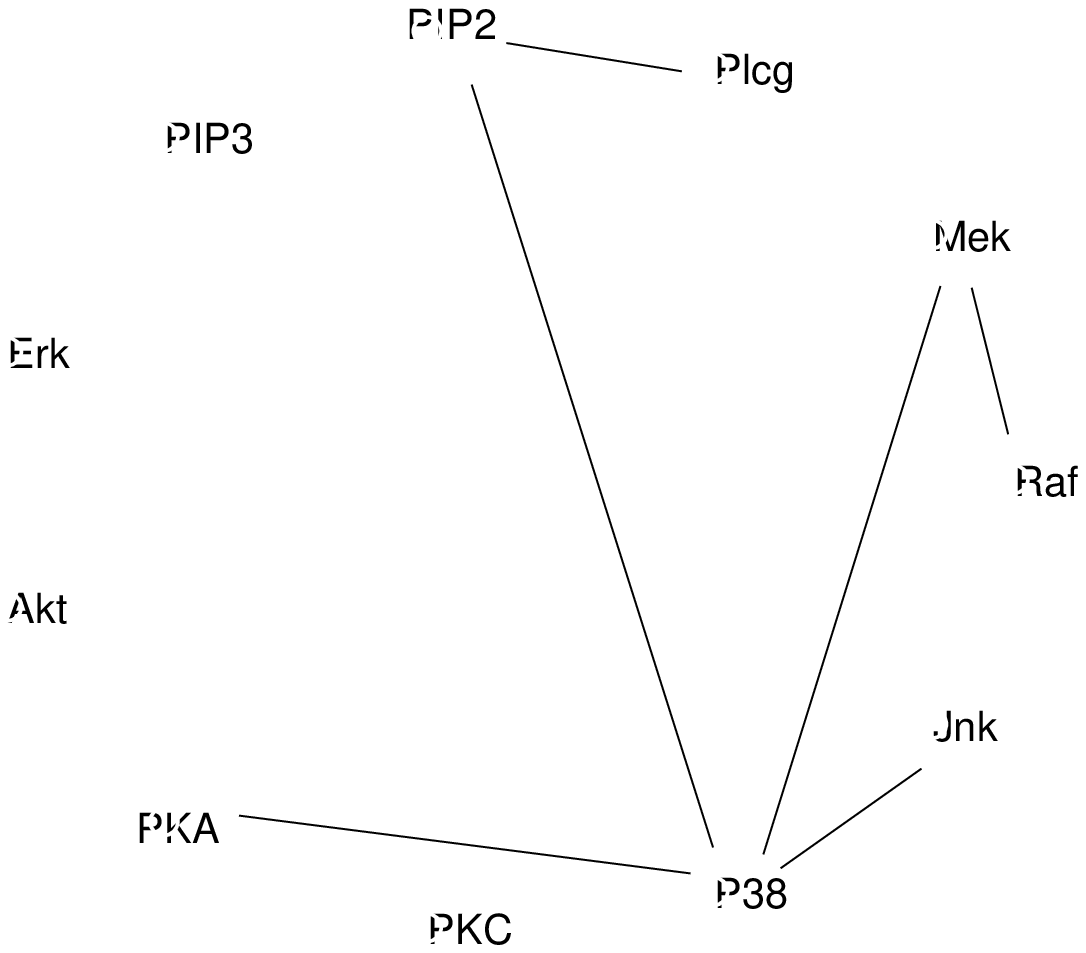}\includegraphics[scale=0.31]{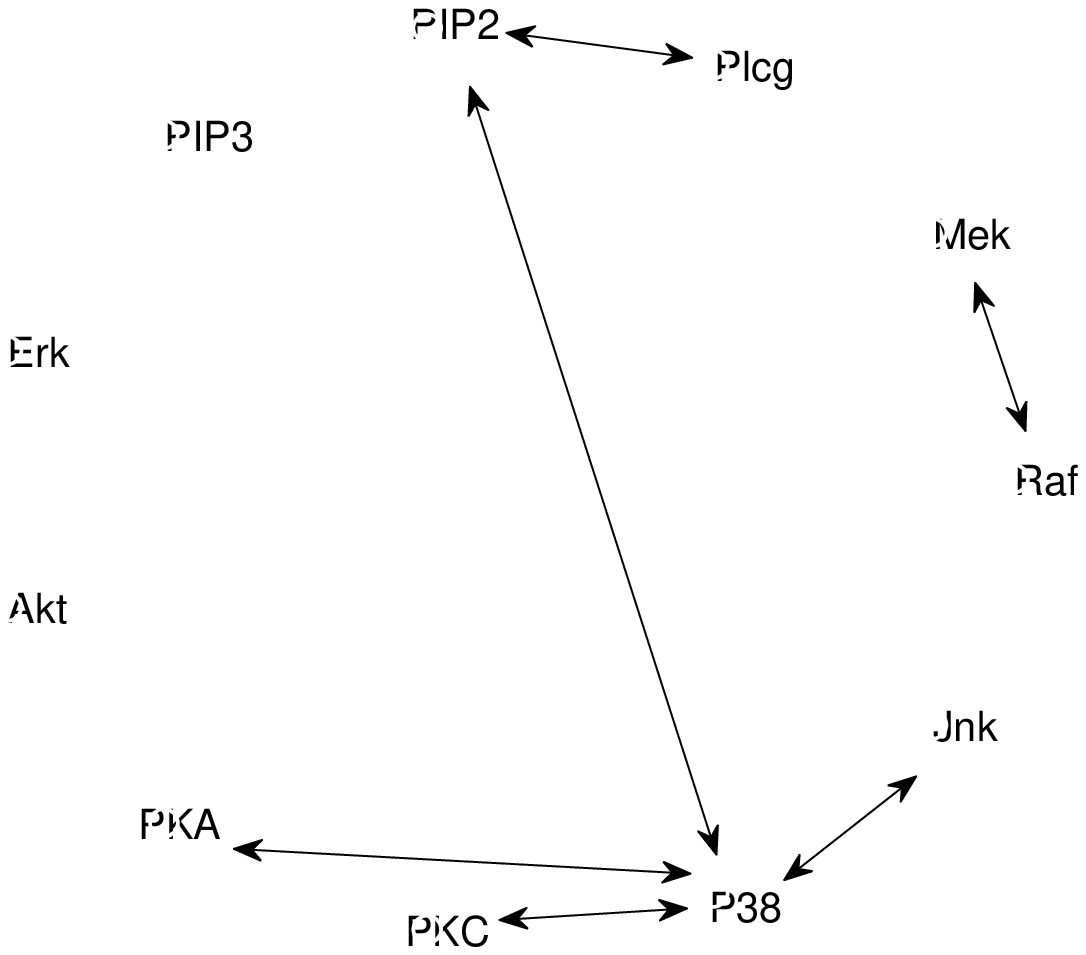}\includegraphics[scale=0.31]{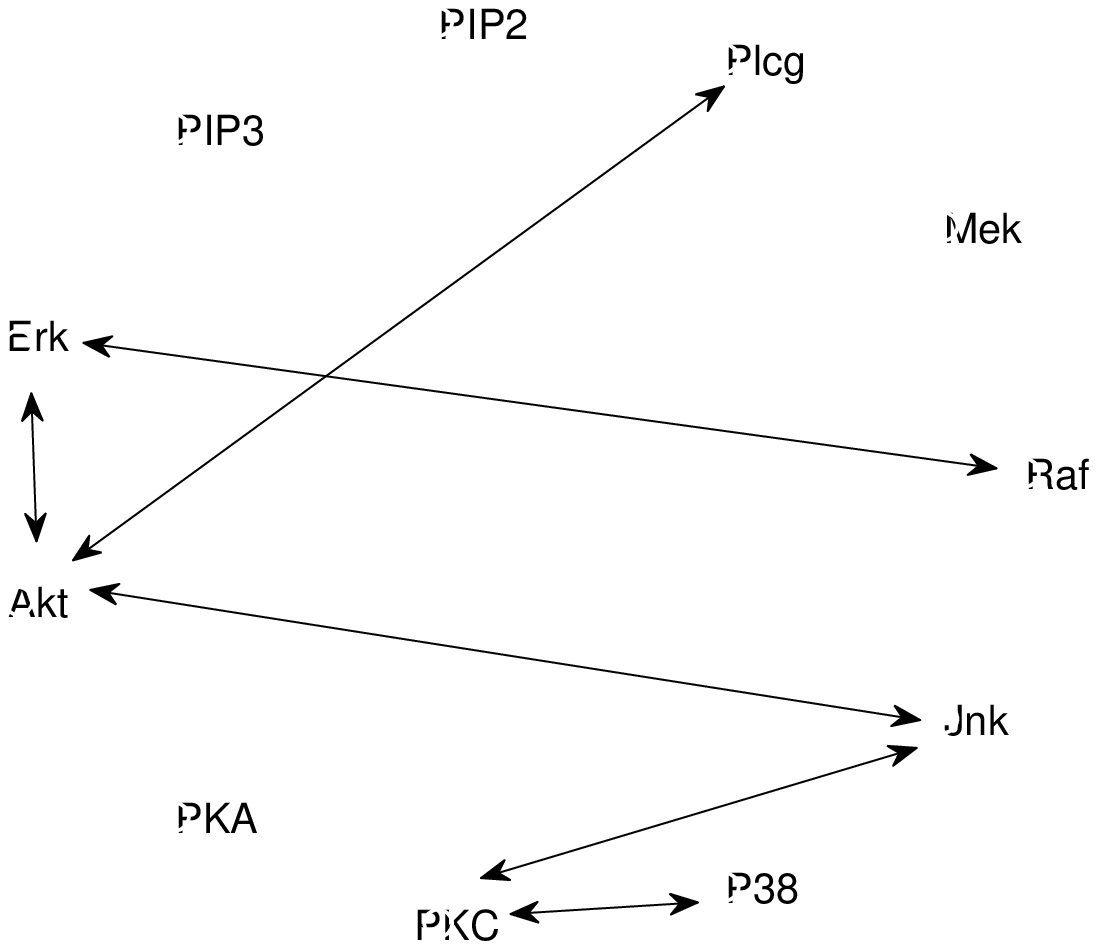}}\tabularnewline
(a) & (b) & (c) & \multicolumn{2}{c}{(d)}\tabularnewline
\end{tabular}
\par\end{centering}
\caption{\label{fig:9-1} (a) Estimated covariance graphs obtained using $\textrm{BCD}$;
(b) Markov graphs obtained using the Graphical Lasso for precision
matrix estimation; (c) Estimated covariance graphs obtained using
the method of \cite{xu2022proximal}; (d) Estimated covariance graphs
obtained using the method of \cite{bien2011sparse}. First row: $k=16$
edges, second row: $k=9$ edges and third row: $k=6$ edges.}
\end{figure*}

\subsubsection{Migration data}

Estimating correlations of migration forecast errors for different
countries helps yielding more accurate estimates and projections of
international migration, which is important for policy making at a
country level. The 2012 revision of the international migration data
(as used in \cite{xu2022proximal}) from the United Nations World
Population Prospects division consist of net migration rate estimates
in each country every five years starting from 1950 ($12$ measurements).
Following \cite{xu2022proximal}, we determine the residual errors
for net migration between all countries using an AR(1) model that
yields $n=11$ observed samples, and estimate the covariance matrix
for $p=5$ countries chosen from the total of $191$ countries. From
the estimated covariance matrix, we determine the correlation across
the chosen countries. For the purpose of illustration, we first choose
the following five countries: Estonia, Latvia, Lithuania (northern
Europe region), Vietnam (South-eastern Asia), and Hong Kong (Eastern
Asia). Fig. \ref{fig:7}(a) shows the heatmap of the Pearson correlation
estimates obtained using SCM for this set of countries. In Fig. \ref{fig:7}(b)
we show the variation of EBIC with $\alpha$ for this example and
find that the recommended $\alpha$ lies anywhere between $0.002$
and $0.07$. Using $\alpha=0.002$ we obtain the correlation matrix
with BCD and show it in Fig \ref{fig:7}(c). The non-zero values of
the correlation estimates between the country pairs Latvia-Estonia
and Latvia-Lithuania is probably due to the fact that Latvia shares
borders with both Estonia and Lithuania. We also consider a second
set of five countries: Bahamas, Puerto Rico, Jamaica (Caribbean),
Zambia and Zimbabwe (Eastern Africa). Figs. \ref{fig:7}(d)-(f) show
their Pearson correlation estimates, the variation of EBIC with $\alpha$,
and the correlation estimates obtained via BCD with $\alpha=0.065$.
From Fig. \ref{fig:7}(f), we observe that the non-zero entries estimated
with BCD correspond to the country pairs that belong to the same region
of the UN world partition. Additionally it is found that Puerto Rico
is uncorrelated to Jamaica and Bahamas, even though they are in the
same region of the UN world partition. This can mainly be attributed
to Puerto Rico being an unincorporated territory of the United States,
whereas Jamaica and Bahamas being independent nations. Moreover, the
official language of Puerto Rico is Spanish and English, with Spanish
being the predominant language in the island. On the other hand, both
Jamaica and Bahamas have English as their official language, with
most people speaking English or English-based creoles.

\subsubsection{Cell signalling data}

The cell signalling dataset \cite{sachs2005causal} has been previously
studied in the context of sparse covariance estimation by the authors
of \cite{bien2011sparse} and \cite{xu2022proximal}. The data contain
the flow cytometry measurements of $p=11$ proteins in $n=7466$ cells.
A missing edge between two proteins in the covariance graph suggests
that the concentration of one protein gives no information about the
concentration of the other protein. 

We estimate the covariance matrix with different sparsity levels for
the flow cytometry measurements using $\textrm{BCD}$ (different sparsity
levels in the covariance matrix are obtained using different values
of $\alpha$). The corresponding covariance graphs with $k=16,9$
and $6$ edges are shown in the first column of Fig. \ref{fig:9-1}.
We compare these estimated covariance graphs with the Markov graphs
(with the same number of edges), as shown in the second column of
Fig. \ref{fig:9-1}, which are undirected graphs obtained by solving
the graphical Lasso problem for precision matrix estimation. Since
a missing edge in the Markov graph indicates conditional independence
between two variables, as opposed to marginal independence in the
case of a covariance graph, the two graphs are not completely identical.
However, as shown in the figure, the covariance graph obtained via
BCD is quite similar to the Markov graph, and this similarity increases
as the number of edges decreases (with completely identical graphs
when $k=9$ and $6$). The third and the fourth column of the figure
show the covariance graphs obtained using the methods in \cite{xu2022proximal}
and \cite{bien2011sparse}. The covariance graphs using the method
of \cite{xu2022proximal} is also similar to the Markov graph (the
covariance graphs obtained with $\textrm{BCD}$ have a larger number
of edges in common with the Markov graphs), whereas the covariance
graphs obtained using the method of \cite{bien2011sparse} are quite
different from the Markov graphs.
\begin{figure}[h]
\begin{centering}
\begin{tabular}{c}
\includegraphics[scale=0.4]{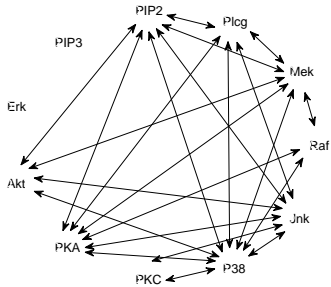}\tabularnewline
\end{tabular}
\par\end{centering}
\caption{\label{fig:10}Estimated covariance graph using BCD with the value
of $\alpha$ ($0.004)$ obtained via EBIC.}
\end{figure}

We also estimate the covariance matrix using BCD with $\alpha$ obtained
via EBIC, the covariance graph of which is shown in Fig. \ref{fig:10}.
This estimated covariance graph has a sparsity of $58\%$ and $23$
edges. 

\section{Conclusions}

In this paper we have presented two methods for estimating sparse
covariance matrices. The proposed methods estimate the sparsity pattern
of the target covariance matrix using FDR multiple hypothesis testing
before solving the MLE problem via two different algorithms. The first
block coordinate descent method does not require the tuning of any
hyper-parameter, whereas the second proximal distance method is computationally
fast but requires the careful tuning of a hyper-parameter. We tested
the proposed algorithms on both synthetically generated and real-world
data, and compared them with two state-of-the-art methods. 

\bibliographystyle{ieeetr}
\bibliography{reference_sparse}

\end{document}